\documentclass[aps,pra,twocolumn,superscriptaddress]{revtex4-1} 

\usepackage[T1]{fontenc} 
\usepackage{amssymb}
\newcommand{\br}{\mathbf{r}} 
\newcommand{\zb}{\bar{z}}

\usepackage{hyperref} 
\hypersetup{
    bookmarks=true,         
    unicode=true,          
    pdftoolbar=false,        
    pdfmenubar=true,        
    pdffitwindow=true,      
    pdftitle={My title},    
    pdfauthor={Author},     
    pdfsubject={Subject},   
    pdfnewwindow=true,      
    pdfkeywords={keywords}, 
    colorlinks=true,       
    linkcolor=red,          
    citecolor=blue,        
    filecolor=blue,      
    urlcolor=blue          
} 

\usepackage[all]{hypcap} 
\usepackage{graphicx}
\usepackage{amsmath}
\usepackage{natbib}
\usepackage{layouts}
\usepackage{verbatim}

\usepackage{mathtools}

\DeclarePairedDelimiter\floor{\lfloor}{\rfloor}

\begin{document} 

\title{Semi-Local Parameterization of the Electron Localization Function\\ in Second-Order Density Gradients}

\author{Alexander Lindmaa} 
\email{alexander.lindmaa@mail.huji.ac.il} 
\affiliation{Racah Institute of Physics, The Hebrew University of Jerusalem, Jerusalem 91904, Givat Ram, Israel} 
\author{Joel Davidsson}
\email{joel.davidsson@liu.se}
\affiliation{Department of Physics, Chemistry and Biology (IFM), Link\"oping University, SE-581 83  Link\"oping, Sweden} 
\author{Ann E.~Mattsson} 
\email{aematts@lanl.gov} 
\affiliation{Los Alamos National Laboratory, XCP-5, P.O.~Box 1663, MS F644, Los Alamos, New Mexico, USA} 
\author{Rickard Armiento}
\email{rickard.armiento@liu.se} 
\affiliation{Department of Physics, Chemistry and Biology (IFM), Link\"oping University, SE-581 83 Link\"oping, Sweden} 
\date{\today} 

\begin{abstract} 
The electron localization function (ELF) is a universal measure of electron localization that allows for, e.g., an effective characterization of physical bonds in molecular and solid state systems. In the context of the widely used Kohn-Sham density-functional theory (KS-DFT) and its generalizations, ELF is given in terms of the single-particle electron density as well as the non-interacting kinetic energy density (KED) of the KS system. Starting from the notion of an edge electron gas put forth by Kohn and Mattsson, we here use an \emph{exactly soluble}, strongly correlated few-electron model of a harmonically confined electron gas in order to parameterize the positive-definite non-interacting KS KED in terms of the density and its reduced second-order gradients. We arrive at a simple, yet generally applicable functional approximation to ELF expressed in the electron density and its derivatives. To demonstrate the validity of our approach, we use the obtained parameterization to perform topological analysis of the bonds in solid Al and Si, and to study physical and chemical adsorption of graphene on a Ni surface. We find that while the expression does not provide a quantitatively accurate approximation of absolute ELF values, the most essential qualitative features are captured. Hence, the expression is useful in contexts where the electron density is available, but not the KS orbitals or the KED, and one desires a qualitative picture of the electron localization that mimics ELF. 
\end{abstract} 

\maketitle 

\section{Introduction} 
Electron localization is an elusive quantum phenomenon that can be quantified in terms of the pair density or Fermi hole, i.e., the conditional probability of observing an electron near a spatial point $\mathbf{r}$ when an electron of the same spin has already been observed at $\mathbf{r}$ \cite{lowdin_2, doi:10.1063/1.460619}. Its effects has been a relevant subject of study since the early attempts at the classification of chemical bonds \cite{pauling_bond,lewis} and remains a topic of fundamental importance in the physics of molecules and solids (see, e.g., Refs.~\onlinecite{PhysRevB.90.241107, PhysRevB.77.115108,doi:10.1021/jp9820774} and references therein). In contrast, the electron localization function (ELF) is a universal, unitless measure of the degree of the electron localization that depends on a single spatial coordinate and which was proposed by Becke and Edgecombe \cite{doi:10.1063/1.458517}. While the original definition of ELF is based on Hartree-Fock (HF) theory, it was later successfully used by Savin \emph{et al.} \cite{doi:10.1002/anie.199201871,Silvi:1994aa} to perform electronic structure calculations in the context of Kohn-Sham density-functional theory \cite{hohenberg_inhomogeneous_1964,kohn_self-consistent_1965}. This was essentially achieved \cite{doi:10.1002/anie.199201871,doi:10.1021/jp992783k} by formally identifying the Laplacian of the conditional probability (i.e., the curvature of the Fermi hole) obtained from a single determinental HF wave function, with the electron- and non-interacting kinetic energy densities (KED) as is given by KS DFT.  Expressed in this way, ELF has, e.g., been widely applied in the context of characterizing chemical bonds in molecular and solid state systems \cite{doi:10.1002/anie.199718081,gatti_2005}. In passing, a pair density functional, dubbed the electron localizability indicator has also been proposed \cite{doi:10.1002/qua.10768}. 

In KS DFT all physical quantities are in principle universally derivable from the electron density, and so it is of fundamental relevance to construct functionals that allows one to investigate bonds in a system directly from the electron density, i.e., without the explicit dependency on the KED. Such expressions are useful in practice when, e.g., using DFT-based methods that do not produce the KED such as orbital-free DFT \cite{parr_yang_1989} (popularly used with pseudopotentials adopted to approximate KS-DFT results) and when analyzing results that have been made available in a database that provides the electron density but not necessarily the KED. This is a common setting for one of the software packages most frequently used for such databases, i.e., the Vienna Ab Initio Software Package (VASP) \cite{PhysRevB.48.13115,PhysRevB.54.11169,PhysRevB.59.1758}. Previous works have extensively discussed various indicators of electron shells and bonds formulated solely in the electron density and its higher-order derivatives. These range from straightforward suggestions of, e.g., simply plotting the reduced gradient, or, the Laplacian; to more intricate constructs such as charge partition schemes \cite{bader1994atoms}. However, these expressions are generally not put on a form that directly mimics ELF itself. 

This paper builds on a series of works that utilize model systems to describe many-electron physics in KS DFT  \cite{armiento_subsystem_2002,armiento_alternative_2003, armiento_functional_2005, hao_subsystem_2010, mattsson_armiento_2010,PhysRevB.90.075139,doi:10.1063/1.4871738}. The concept of a model system goes back to the notion of an edge electron gas introduced by Kohn and Mattsson \cite{kohn_edge_1998}, and to the original construction of the local density approximation (LDA) of the exchange energy based on the uniform electron gas (UEG) \cite{fermi_metodo_1927,thomas_calculation_1927,dirac_note_1930,hohenberg_inhomogeneous_1964}. Hence, we take the harmonic oscillator (HO) model system to be a harmonically confined, non-interacting fermion gas, which captures strong electron correlation by confining the electrons in one dimension. By solving the KS equations that describe this system we construct a parameterization of the non-interacting positive-definite KED which is then used to approximate ELF. The result is a semi-local functional expression in the electron density which, by design, should reflect the most salient features of ELF in real, \emph{interacting}, electronic systems at least in regions that are well described by the HO gas regime. A major objective of this work is hence to investigate to what extent the HO model, as it stands, is capable of describing electron localization in more realistic situations. To achieve this we analyze the transferability of the expression by comparing it to the real ELF (i.e., calculated using the KED from the KS orbitals) in example systems of covalent, metallic, and dispersive bonds. It is found that our parameterization qualitatively reproduces many essential features of ELF in regions where it takes on moderate to high values (viz.~high localization), but it cannot in general be used as a direct quantitative approximation of ELF.  

The rest of the paper is organized as follows. Section~II briefly presents the central equations related to ELF that are of relevance to our discussion. In Sec.~III we restate some of the formal properties of the non-interacting KED in the HO model system. In Sec.~IV we put forward the main result of the present work: the derivation of the parameterization of ELF based on the HO model system of confinement. Sec.~V is devoted to our results, i.e., (\emph{i}) the evaluation of the parameterization of ELF along bonds in solid Al and Si, as well as on two-dimensional topological surfaces, and (\emph{ii}) the case of physical and chemical adsorption of graphene on a Ni surface. Finally in Sec.~VI we present our summary and conclusions. 

\section{The Electron localization function} 
We begin by considering a system with a spin-polarized electron density $n_{\sigma}(\mathbf{r})$, where $\sigma$ is the spin degree of freedom (Hartree atomic units are used throughout). In KS DFT, the density is represented by 
\begin{equation}\label{eq:density}
	n_{\sigma}(\br) = \sum_{\nu:\varepsilon_{\nu} \leq \mu} |\phi_{\sigma \nu}(\br)|^2, 
\end{equation} 
where $\{ \phi_{\sigma\nu} \}^{\infty}_{\nu=1}$ are KS singe-particle orbitals with eigenvalues $\varepsilon_{\nu}$, and $\mu$ is the self-consistent chemical potential (In the following, we let $\mu = 0$). It is convenient to work with expressions that are invariant under density-scaling, and thus we adopt in this work the standard notion of a generalized gradient approximation \cite{perdew_accurate_1986}. Within this framework, the reduced density gradient $s$ and Laplacian $q$ are scale invariant, dimensionless quantities that depend on the density according to 
\begin{eqnarray} 
s = \frac{|\nabla n|}{2(3\pi^2)^{1/3}n^{4/3}} \label{eq:s} \\
q = \frac{\nabla^2n}{4(3\pi^2)^{2/3}n^{5/3}} \label{eq:q}, 
\end{eqnarray}
from which it follows that $s^2 \sim q$. Of great importance to the following discussion will be the ELF probability index. Following the notation of previous works \cite{hao_subsystem_2010}, we define ELF as 
\begin{equation}\label{eq:ELF} 
\mathrm{ELF}=\frac{1}{1+(D/D_h)^2} . 
\end{equation}  
where
\begin{equation}\label{eq:ddh_def}
\frac{D}{D_h}(\br)=\frac{\tau_{\sigma}(\br) - \tau_{\mathrm{W}}(\br)}{\tau_{\mathrm{TF}}(\br)}.
\end{equation} 
Here, $\tau_{\sigma}(\br)$ is the non-interacting positive-definite kinetic energy density (KED), expressed in the KS orbitals as 
\begin{equation}\label{eq:tau_pos}
\tau_{\sigma}(\br)=\sum_{\nu:\varepsilon_{\nu} \leq \mu}|\nabla \phi_{\sigma \nu}(\br)|^2 .
\end{equation} 
As a final prerequisite for our considerations, we recall the Thomas-Fermi (TF) KED $\tau_{\mathrm{TF}}(\br)$, which is given by   
\begin{equation} 
\tau_{\mathrm{TF}}(\br) = \frac{3}{10}(2\pi^2)^{2/3}\,n_{\sigma}^{5/3}(\br),
\end{equation} 
as well as 
\begin{equation}\label{eq:von_weiz} 
\tau_{\mathrm{W}}(\br) = \frac{1}{8} \frac{|\nabla n_{\sigma}(\br)|^2}{n_{\sigma}(\br)}, 
\end{equation}
which is the von Weizs\"acker or single-particle KED. Importantly, the difference 
\begin{equation}\label{eq:pauli}
\tau_{\mathrm{p}}(\br) \equiv \tau_{\sigma}(\br)-\tau_{\mathrm{W}}(\br)
\end{equation}
in the numerator of Eq.~(\ref{eq:ddh_def}) is commonly referred to as the Pauli KED, since it is a point wise measure of the Pauli exclusion principle and hence the fermionic character of the KED. The von Weizs\"{a}cker term, $\tau_{\mathrm{W}}$, of Eq.~(\ref{eq:von_weiz}) is then being interpreted as the KED of a bosonic system with the same density \cite{Silvi:1994aa,doi:10.1063/1.4871738}. To see this, consider a bosonic wave-function $\Psi_{\mathrm{b}}(\br)$ defined by 
\begin{equation}
	\Psi_{\mathrm{b}}(\br) = \sqrt{\frac{n(\br)}{2N}}, 
\end{equation}
where $N$ is the total number of orbitals in the system, allowing $2N$ bosons reside in the same orbital, and using Eq.~(\ref{eq:tau_pos}), we see that 
\begin{equation}
\tau_{\mathrm{b}}(\br) = N\left| \nabla \sqrt{\frac{n(\br)}{2N}} \right|^2 = \frac{1}{2}\left| \frac{\nabla n(\br)}{2 \sqrt{n(\br)}} \right|^2 = \frac{1}{8} \frac{|\nabla n_{}(\br)|^2}{n_{}(\br)}, 
\end{equation}
thus retrieving Eq.~(\ref{eq:von_weiz}). As such, the von Weizs\"{a}cker term provides a lower bound to the kinetic energy of a fermionic state \cite{lieb_thomas-fermi_1973}, and thus, as a result of Eq.~(\ref{eq:pauli}) describes a highly localized fermion.

By normalizing with the TF KED in Eq.~(\ref{eq:ddh_def}), the value of ELF = 1/2 corresponds to the UEG that the TF KED was originally derived from. ELF approaches zero only when the kinetic energy goes to infinity which happens for fermions far inside the classically forbidden region, i.e., outside a surface. This limit is the Airy gas (AG) limit discussed at length in earlier works \cite{kohn_edge_1998, armiento_functional_2005, PhysRevB.90.075139}. Finally, for highly localized fermions ELF $\to 1$, which corresponds to the limit of perfect confinement, and which we attempt to address in the present article.  


\section{The Harmonic oscillator model system of confinement} 
The formal properties of the harmonic oscillator (HO) electron gas has been detailed by some of us in previous works \cite{armiento_subsystem_2002,PhysRevB.90.075139}. For the purposes of this work, we begin by restating the most central equations. Thus, in the parlance of previously mentioned works, we take the HO model system to mean a KS DFT system where the effective potential is constant in the two spatial directions $x$ and $y$, but confines the KS states along the $z$-direction according to the familiar linear harmonic oscillator. Thus, we let 
\begin{equation} 
v^{\mathrm{HO}}_{\mathrm{KS}}(\mathbf{r}) = \frac{\omega^2}{2} z^2,
\end{equation}
where $\omega$ is the angular frequency of the harmonic oscillator. Solving the corresponding KS equations along the $z$-axis results in orbitals of the form 
\begin{eqnarray}\label{eq:ho_orb} 
l^{1/2}\phi(\zb) &=& \tilde{\phi}(\zb) \label{eq:phi}, \\
\tilde{\phi}_j(\zb) &=& \left( \frac{1}{\sqrt{\pi}2^{j}j!} \right)^{1/2}H_j(\zb)e^{-\zb^2}\label{eq:tilde_phi}, 
\end{eqnarray} 
where $\{H_{j}\}^{\infty}_{j=0}$ are Hermite polynomials, and the corresponding eigenvalues are given by 
\begin{equation} 
\varepsilon_j = \frac{1}{l^2} \left(j + \frac{1}{2} \right), \quad j = 0, 1, \ldots .
\end{equation} 
Here we have defined the surface thickness $l = \sqrt{1/\omega}$ of the edge region \cite{kohn_edge_1998}, so that $\bar{z} = z/l$. In addition we define 
\begin{equation}
  \quad w= \frac{\omega}{\mu}, \quad \alpha = \frac{1}{w} - \frac{1}{2}, \quad N(\mu) = \floor{\alpha}, 
\end{equation}
where $w$ is a curvature parameter of the potential parabola, $N(\mu)$ is the index of the highest occupied orbital in the $z$-dimension and $\mu$ is given by 
\begin{equation}\label{eq:chem_pot}
\mu = \left( \alpha + \frac{1}{2}\right)\frac{1}{l^2} .  
\end{equation}
Eq.~(\ref{eq:chem_pot}) thus defines the \emph{degree of fermionic confinement} $\alpha$ which, by virtue of being inversely proportional to $w$, denotes the index of the highest occupied KS state $z$-direction. Consequently, the remainder, $0 \leq \alpha-N < 1 $, refers to the continuous bands in the $xy$-directions. From Eq.~(\ref{eq:ho_orb}) the single-particle density and the positive KED can now be expressed exactly in terms of the variables $\alpha$ and $\bar{z}$ according to \cite{doi:10.1063/1.4871738}  
\begin{equation}\label{eq:dimless_dens} 
l^3n(\zb) = \frac{1}{\pi^{3/2}} \sum^{N}_{j = 0}\frac{1}{2^j2!}H_j^2(\zb)e^{-\zb^2}(\alpha - j) 
\end{equation} 
and  
\begin{equation}\label{eq:dimless_tau} 
l^5\tau(\zb) = \frac{1}{2 \pi}\sum^{N}_{j=0}\left[ (\alpha -j)^2 \tilde{\phi}^2_j(\zb) + (\alpha -j)\tilde{\phi}^{2\prime}_j(\zb)\right],
\end{equation}
where $n(\bar{z})$ and $\tau({\bar{z}})$ are dimensionless by construction. Here we make an additional crucial point. While Eqs.~(\ref{eq:dimless_dens})--(\ref{eq:dimless_tau}) are valid for any value of $\alpha$, e.g., a large one which corresponds to a wide harmonic potential, we are concerned with the opposite, few-electron limit. We take this limit, which corresponds to a narrow harmonic potential where the KS states are highly confined, to be starting point for our further considerations. For a detailed discussion regarding the many-particle limit, we refer to Ref.~\onlinecite{PhysRevB.90.075139}. In the next Section, we will derive a parameterized version of ELF as defined by Eq.~(\ref{eq:ddh_def}), via the expression for $D/D_h$ of Eq.~(\ref{eq:ddh_def}). The resulting parameterization will thus be given in terms of the reduced density gradient $s$ and Laplacian $q$. 


\section{Parameterization of the KED in the HO model system} 

We begin this section by deriving an exact expression for $D/D_h$ of Eq.~(\ref{eq:ELF}) in terms of $s$ and $q$, for $\alpha$'s in the range $(0,1]$. At this point it is stressed that for $\alpha > 1$, which corresponds to a higher number of occupied states, the sheer complexity of the expressions given by the Eqs.~(\ref{eq:dimless_dens})--(\ref{eq:dimless_tau}) becomes prohibitively difficult to handle. It is de facto highly non-trivial to exactly map $\alpha$ and $\bar{z}$ to $s$ and $q$ in these cases. In order to deal with this issue, we will instead resort to the exact parameterization of $D/D_h$ in the range $(0,1]$ in order to construct an extended parameterization which is based on $\alpha$'s in the range $(0,2]$. In the present article, we work under the assumption that such an extended, parameterized version of $D/D_h$ should be able to constitute an approximation that can be applied also in the case for higher values, i.e., $\alpha > 2$. We expand further on this points later in this section. 

Continuing with the exact parameterization, the reduced gradient and Laplacian of Eqs.~(\ref{eq:s})--(\ref{eq:q}) are straightforward to derive from any given exact HO model density. Thus, for $\alpha \in (0,1]$ and for any $\bar{z}$ one can verify that 
\begin{equation}
\begin{array}{l}
\displaystyle s(\alpha,\bar{z}) = \frac{|2\bar{z}|}{2(3\sqrt{\pi}\,\alpha\,e^{-\bar{z}^2})^{1/3}},\\
\\
\displaystyle q(\alpha,\bar{z}) = \frac{2 \bar{z}-1}{2(3\sqrt{\pi}\,\alpha \, e^{-\bar{z}^2})^{2/3}}, 
\end{array} 
\label{eq:sq_rels} 
\end{equation}
and for $\alpha \in [1,2]$, and $z \in (-\infty,\infty)$, 
\begin{equation}
\begin{array}{l}
\displaystyle s(\alpha,\bar{z}) = \frac{|(\alpha-2)\bar{z}-2(\alpha-1)\bar{z}^3|}{(3\sqrt{\pi}e^{-\bar{z}^2})^{1/3}[\alpha+(2\alpha-1)\bar{z}^2]^{4/3}},\\ 
\\ 
\displaystyle q(\alpha,\bar{z}) = \frac{2(\alpha-2)\bar{z}^4+(-4\alpha+5)\bar{z}^2+\frac{\alpha}{2}-1}{(3\sqrt{\pi}e^{-\bar{z}^2})^{2/3}[\alpha+(2\alpha-1)\bar{z}^2]^{5/3}}. 
\end{array} 
\label{eq:sq_rels_2} 
\end{equation}
The expression for $D/D_h$ of Eq.~(\ref{eq:ddh_def}) then becomes 
\begin{equation}\label{eq:ddh1}
\frac{D}{D_h}(\alpha,\bar{z})=\frac{5}{3}\frac{\alpha^{\frac{1}{3}} \,e^{\frac{2\bar{z}^2}{3}}}{(3\sqrt{\pi})^{2/3}}, \quad \alpha \in (0,1], \, \forall \, \bar{z}. 
\end{equation}
We now seek to express $D/D_h$ in the reduced quantities $s(\alpha,\bar{z})$ and $q(\alpha,\bar{z})$ given by the relations of Eq.~(\ref{eq:sq_rels}). By noting that 
\begin{equation}\label{eq:s2q} 
\frac{s^2}{q} = \frac{\bar{z}^2}{\bar{z}^2-\frac{1}{2}} \iff \bar{z}^2=\frac{s^2}{s^2-q}, \quad s^2 \neq q, 
\end{equation} 
and using the second of the relations of Eq.~(\ref{eq:sq_rels}) we get 
\begin{equation}\label{eq:alpha_rel}
\alpha = \left[ \frac{1}{q} \frac{\bar{z}-\frac{1}{2}}{(3\sqrt{\pi})^{\frac{2}{3}}\,e^{-\frac{\bar{z}^2}{3}}} \right]^{\frac{3}{2}}. 
\end{equation} 
Combining Eqs.~(\ref{eq:ddh1}), (\ref{eq:s2q}), and (\ref{eq:alpha_rel}) results in 
\begin{equation}\label{eq:ddh1sq}
{\eta}_1(s,q) \equiv \frac{D}{D_h}(s,q) \bigg|_{\alpha \in (0,1]} = \frac{5}{9\sqrt{2\pi}}\frac{e ^{ \frac{1}{2}\left( \frac{s^2}{s^2-q} \right)}}{\sqrt{s^2-q}} . 
\end{equation} 
The quantity $\eta_1(s,q)$ of Eq.~(\ref{eq:ddh1sq}) defines an \emph{exact} relation for the harmonic oscillator model systems in the range $\alpha \in (0,1]$ and for any value of $\bar{z}$. For such harmonic oscillators, there are no point wise values of $s$ and $q$ such that $s^2-q = 0$, inadvertently showing up as singularities in Eq.~(\ref{eq:ddh1sq}). We note also that the line in the $(s^2,q)$ plane that defines the transition $\alpha: 1 \to 2$ can be written on an exact form which is given by 
\begin{equation}\label{eq:sq_line}
	s^2 = \frac{1}{3}q\left(3 + \frac{1}{W\left(2 (\frac{\pi}{3e})^{1/3}q \right)} \right),
\end{equation}
where $W(\cdot)$ is the Lambert function \cite{stegun_1965}. It is in general not possible to find lines as the one given by Eq.~(\ref{eq:sq_line}) that correspond to additional states being occupied in the harmonic oscillator. 

Keeping the preceding discussion close in mind, we now proceed to discuss the construction of the parameterization of $D/D_h$ for $\alpha \in (0,2]$. The most general form of a parameterized version $\eta(s,q)$ for all such values can be defined in terms of a piecewise function 
\begin{equation}\label{eq:full_expression} 
\eta(s,q) = 
	\begin{cases} 
		\eta_1(s,q),\, \text{when} \, \alpha \in (0,1] \\ 
		\eta_2(s,q),\, \text{when} \, \alpha \in [1,2] . 
	\end{cases}
\end{equation} 
However, as previously discussed it is highly nontrivial to construct a closed form expression of $\eta_2(s,q)$ in a similar fashion as was done for the case of $\eta_1(s,q)$. It nevertheless seems plausible that $\eta_1(s,q)$ should contain some transferable features, that would also be present in an expression for $\eta_2(s,q)$, even though they describe different aspects of the HO model. Framed differently, the main attributes that are contained in $\eta_1(s,q)$ should contain a set of general features, that would resemble the exact ELF also for values of $\alpha \in [1,2]$. With this in mind, we consider a modified expression written in the form 
\begin{equation}\label{eq:finaleq}
\tilde{\eta}(s,q) = \frac{5}{9\sqrt{2\pi}}\frac{e ^{ \frac{1}{2}\left( \frac{\beta s^2}{f_{\epsilon,\delta}(s,q)} \right)}}{[f_{\epsilon,\delta}(s,q)]^{\gamma/2}}, \quad \alpha \in (0,2],
\end{equation} 
where $\beta$, $\gamma$, $\epsilon$, and $\delta$ are real parameters that need to be determined. The properties of the function $f_{\epsilon,\delta}$ will be discussed in the following: since $s^2-q \to 0$, as $\bar{z} \to 0$, and $\alpha \to 2$, it is highly likely that a general density would be such that $s^2-q \leq 0$. Notably, in a real system, there might exist points where $s^2-q \leq 0$, which would lead to unavoidable singularities. To resolve this problem we first replace the problematic occurrences of $s^2-q$ in the denominators with $s^2-q+\epsilon$, where $\epsilon$ is a sufficiently small positive number. Then we introduce a function that approaches $s^2-q+\epsilon$ when that quantity is a large positive value, but rather smoothly approach zero instead of becoming negative when $q > s^2$. Hence, we introduce a function $f_{\epsilon,\delta}$, which we define as 
  \begin{equation}\label{eq:adj} 
f_{\epsilon,\delta}(s,q)=\left(\frac{\epsilon}{\delta}\right) \ln \left[1+e^{\frac{\delta(s^2-q)}{\epsilon}}\left(e^{\delta}-1\right)\right],
\end{equation} 
where $\delta$ sets the sharpness with which $f_{\epsilon,\delta}$ approaches zero. 
Despite the alterations from Eq.~(\ref{eq:ddh1sq}), this remains a very good approximation of the exact $\eta(s,q)$ for HO systems with $\alpha \in (0,1]$. By performing a least-square minimization with respect to the exact ELF for HO models with $\alpha \in (0,2]$ we obtain $\beta=1.122$ and $\gamma=1.420$. In this fit we used $\delta = 10$ and $\epsilon = 0.1$, which results in a good approximation of $\sqrt{s^2-q}$. We again stress that even though $\tilde{\eta}(s,q)$ has been constructed by fitting the values of the parameters $\beta$ and $\gamma$ to the exact ELF on the set $\alpha \in (0,2]$, we take Eq.~(\ref{eq:finaleq}) to be valid values of $\alpha$ that are larger than $2$. 

\begin{figure}[ht]
	\centering 
 	\includegraphics[]{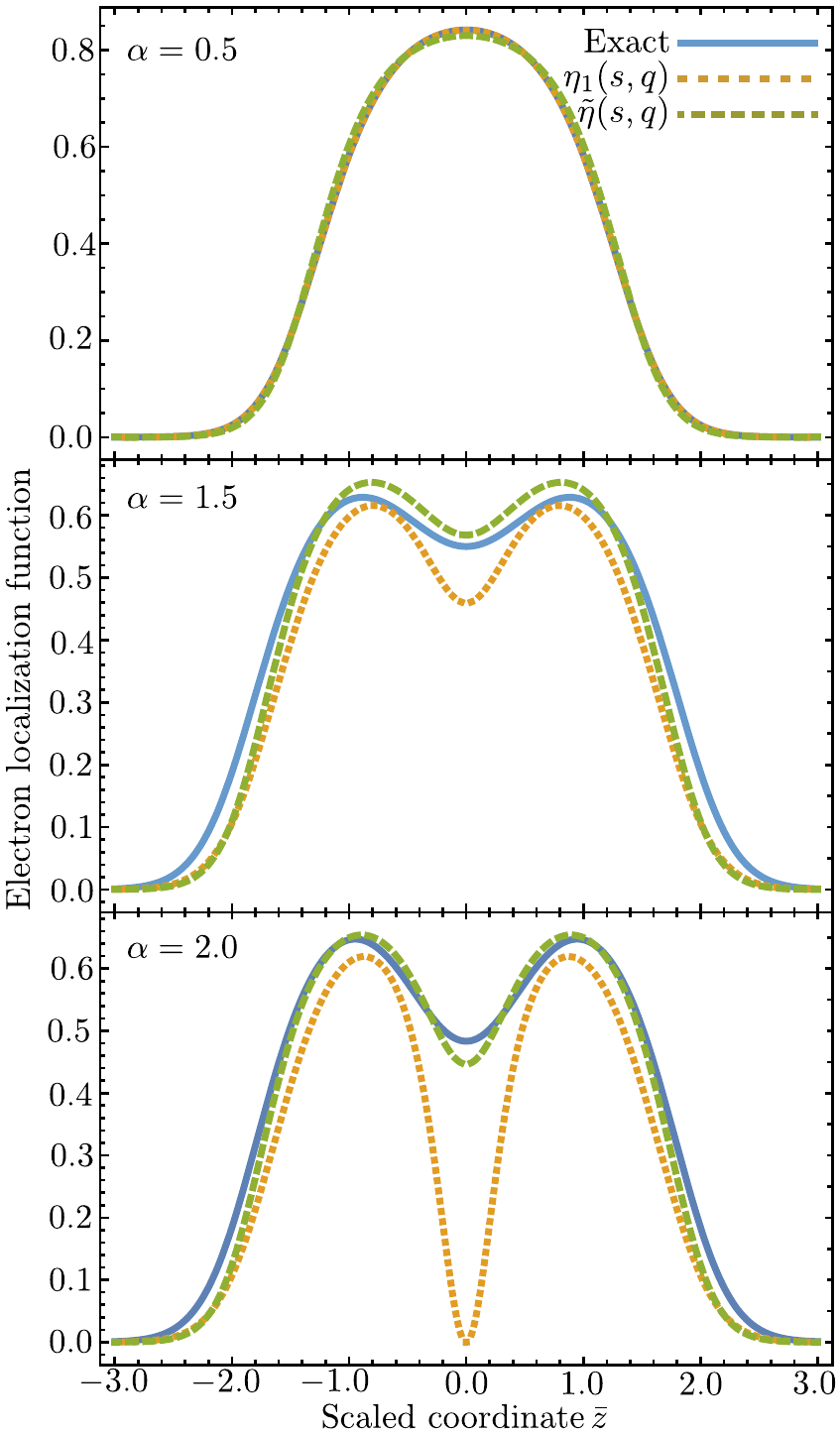} 
  	\caption{The exact ELF for different values of $\alpha$ (number of occupied states) as compared with ELF based on the parameterizations $\eta_1(s,q)$ of Eq.~(\ref{eq:ddh1sq}) and $\tilde{\eta}(s,q)$ of Eq.~(\ref{eq:finaleq}) presented in this work. As the value of $\alpha$ increases (for fixed value of the chemical potential $\mu$) the potential parabola becomes wider, thus representing a less confined system. The top panel shows a highly confined system ($\alpha = 0.5$). The middle panel shows a less confined system ($\alpha =1.5$) where the next state in the oscillator has been occupied. The bottom panel shows a system in which $\alpha =2$.} 
 	\label{fig:fig_1} 
\end{figure}

Fig.~\ref{fig:fig_1} shows the performance of $\eta_1(s,q)$ along with the parameterization $\tilde{\eta}(s,q)$, both in comparison to the exact ELF for different values of $\alpha$ (number of occupied states) in the HO. As is evident from the figure, the parameterized version of ELF which is given by Eq.~(\ref{eq:finaleq}) captures the exact values to a reasonable degree, most importantly in the region around the center of the harmonic oscillator where ELF is larger than $0.5$. We also note that ELF based on $\eta_1(s,q)$ does not deviate significantly from the exact values of ELF for $\alpha \in [1,2]$ (our initial hypothesis), except for the case when $\bar{z} \to 0$ and $\alpha \to 2$. Indeed, as is seen in Fig.~\ref{fig:fig_1} for increasing values of $\alpha$, a sharp dip starts to develop at the symmetry line $\bar{z} = 0$. This reflects the fact that $\eta_1(s,q)$ is unable to fully represent the UEG in this region. This behaviour is further emphasized in the case $\alpha = 1.5$, where two surfaces begin to emerge. In the extreme case $\alpha = 2$, the UEG characteristics that form between the two surfaces can not be restored by $\eta_1(s,q)$. We note however that the dip produced by $\eta_1(s,q)$ at the center of the oscillator is a feature that to some degree carries over to the case $\alpha \in [1,2]$ so as to restore the behaviour of the exact ELF for less confined situations. This situation, in spite of the fact that the HO models defined by the transition $\alpha: 0\to 1$ are starkly different is, by virtue of its construction captured by $\tilde{\eta}(s,q)$, as the dip is significantly suppressed. 

One of the main purposes of this article is to investigate to what extent an ELF based on a one dimensional, harmonically confined electron gas is capable of describing effects of electron localization for more realistic densities. This investigation is the subject of the next section. 


\section{Results and Discussion} 

In the following, the behavior of the ELF parameterization of Eq.~(\ref{eq:finaleq}) is studied for three physical systems with different types of bonding physics: bulk fcc Al (a metal), bulk diamond Si (a semiconductor), and graphene on a Ni surface that allow both chemisorbed and physisorbed bonding depending on the distance between the C and Ni atoms. The values of ELF based on $\tilde{\eta}(s,q)$ are evaluated for $s^2(\br)$ and $q(\br)$ and compared to the real ELF calculated from the KS orbitals in the same system. In order to perform the self-consistent computations, we primarily use the Elk all-electron electronic structure software \cite{code_elk} with the local spin-density approximation \cite{perdew_accurate_1992}. The complete computational details are given in Appendix~\ref{sec:comp_details}. 

\subsection{The parameter $s^2 - q$}

\begin{figure}[ht] 
	\centering 
 	\includegraphics[width=\linewidth]{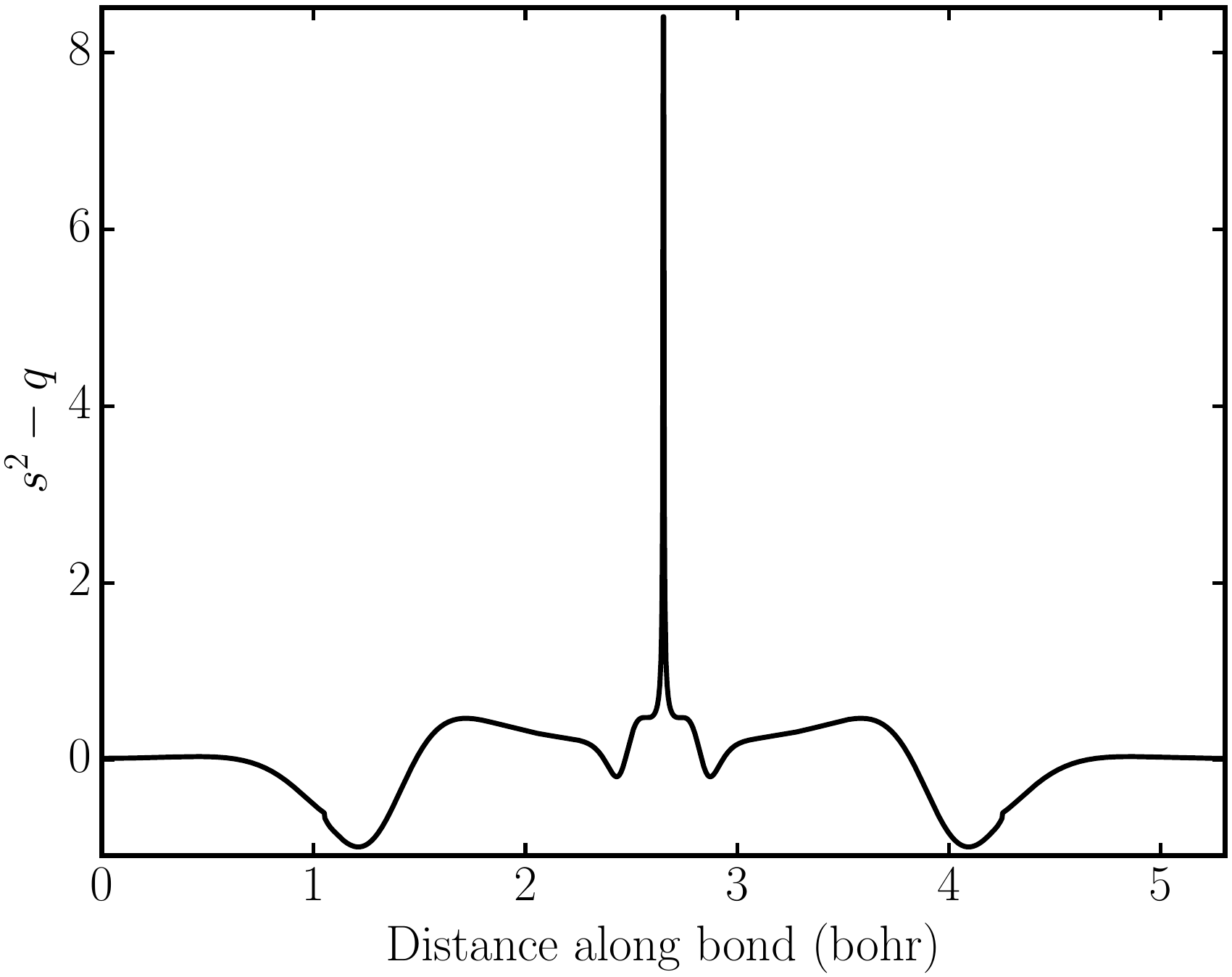} 
  	\caption{The difference $s^2-q$ in solid fcc Al along a nearest-neighbor line between repetitions of individual Al atoms in the unit cell. The sharp peak is located at the Al atom.} 
 	\label{fig:fig_2} 
\end{figure}

\begin{figure}[ht] 
	\centering 
 	\includegraphics[width=\linewidth]{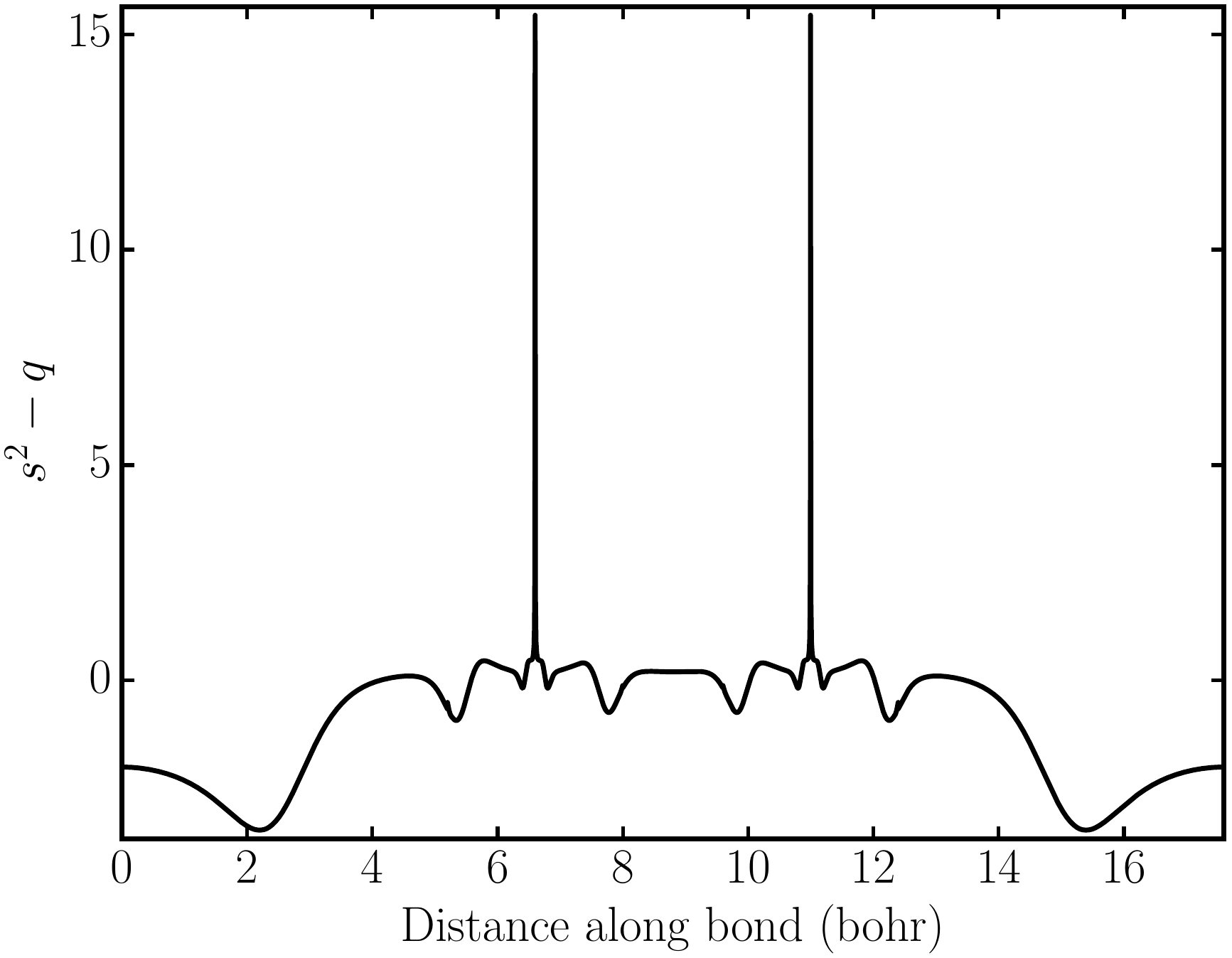} 
  	\caption{The difference $s^2-q$ along a nearest neighbor line in solid Si in the diamond structure. The sharp peaks are located at the two Si atoms in the unit cell, and the line between them intersects the nearest-neighbohr $sp$ - bond. }
 	\label{fig:fig_3} 
\end{figure}

The preceding discussion and derivation makes it clear that $s^2-q$ takes a very central role in the parameterization. This expression is show in Figs.~(\ref{fig:fig_2}) and (\ref{fig:fig_3}) for the Al and Si systems. Both the systems have regions where $s^2-q$ is negative, even though this does not occur in any spatial point in the HO model system. 

\begin{figure}[] 
	\centering 
 	\includegraphics[width=\linewidth]{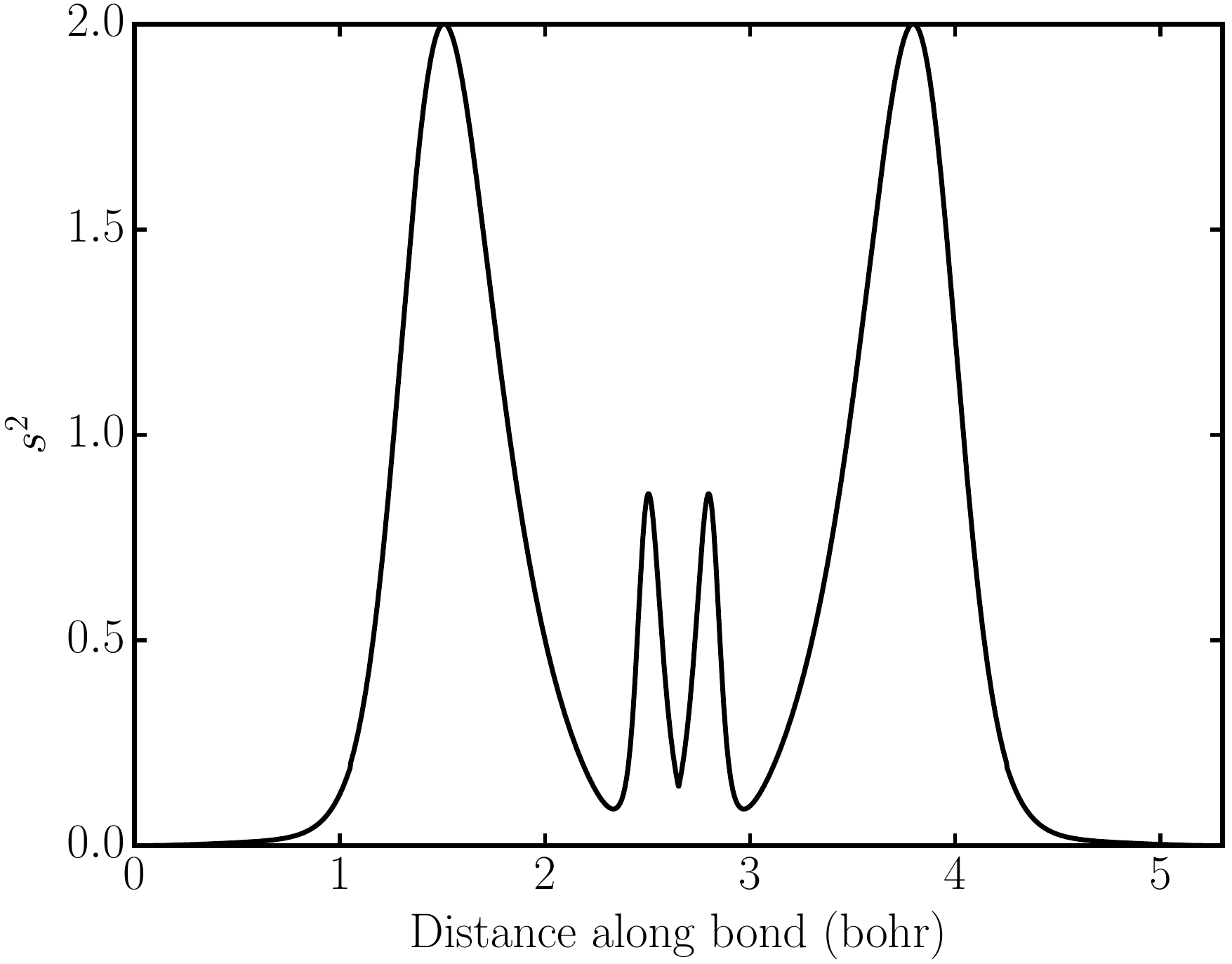} 
  	\caption{The behaviour of $s^2$ in solid fcc Al along a nearest-neighbor line between repetitions of individual Al atoms in the unit cell.}
 	\label{fig:fig_4} 
\end{figure}

\begin{figure}[ht] 
	\centering 
 	\includegraphics[width=\linewidth]{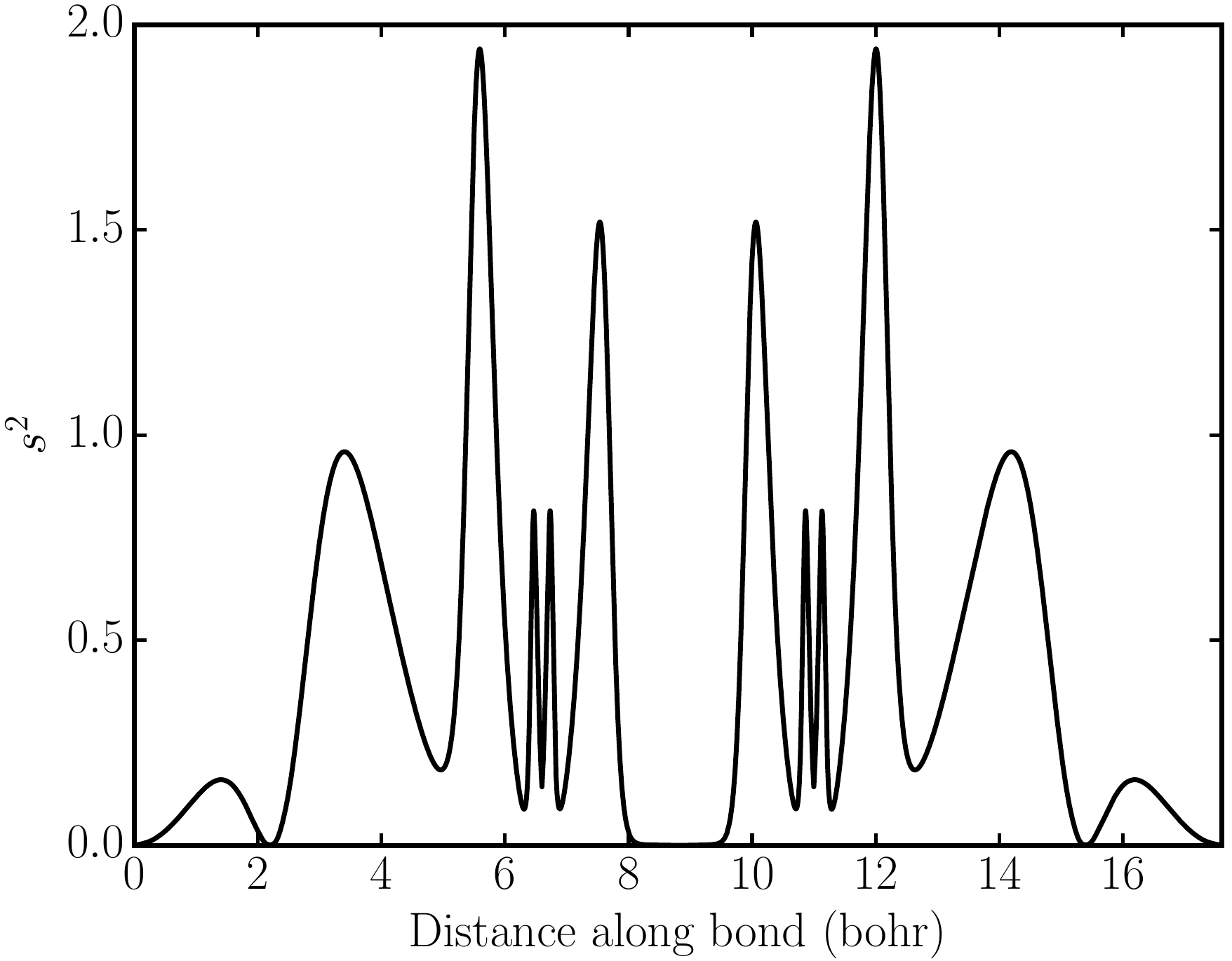}  
  	\caption{The behavior of $s^2$ in solid diamond Si, along a nearest-neighbor line between the two Si atoms in the unit cell.}
 	\label{fig:fig_5} 
\end{figure}

One can think of the $\epsilon$ parameter in Eq.~(\ref{eq:finaleq}) as a slight adjustment upwards of these graphs. This adjustment will reduce the problematic regions of negative $s^2-q$, but not entirely remove them. The adjustment function, Eq.~(\ref{eq:adj}) will make sure the parameterization can be evaluated in all points and still be smooth, but, there is no reason to expect the expression to reproduce the real ELF well in these regions. To further illustrate that $s^2-q$ picks up properties of the underlying system that are not readily seen in other quantities, Figs.~(\ref{fig:fig_4}) and (\ref{fig:fig_5}) show $s^2$ for the Al and Si systems for comparison. 

%
%
%
%

\subsection{Parameterized ELF in Si and Al}

\begin{figure}[ht]
	\centering 
 	\includegraphics[width=\linewidth]{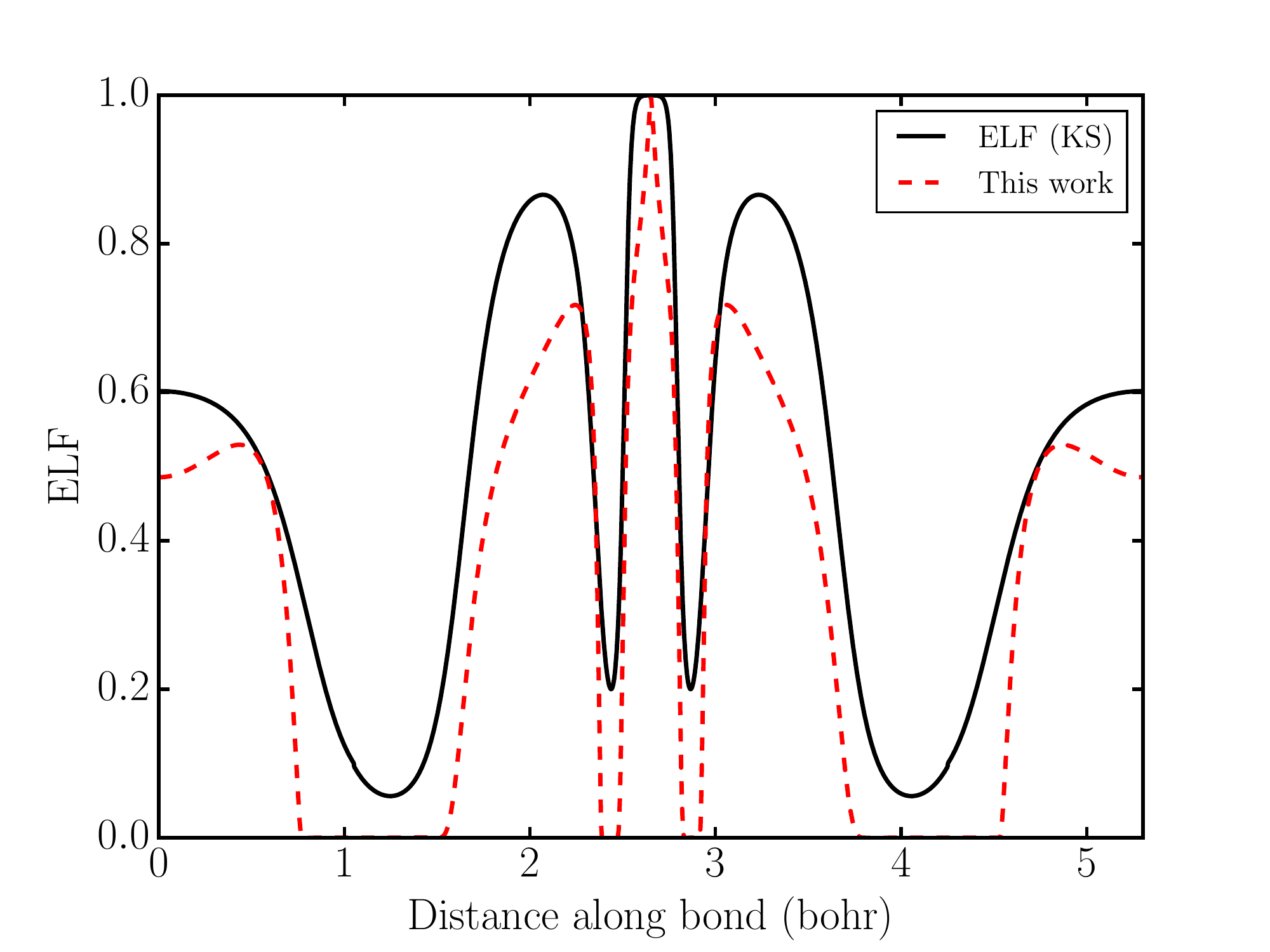} 
   	\caption{(color online) The exact KS DFT ELF (solid, black) and the ELF parameterization in this work (dashed, red) in solid fcc Al along a nearest-neighbor line between repetitions of individual Al atoms in the unit cell.} 
 	\label{fig:fig_6} 
\end{figure}

\begin{figure}[ht] 
	\centering 
 	\includegraphics[width=\linewidth]{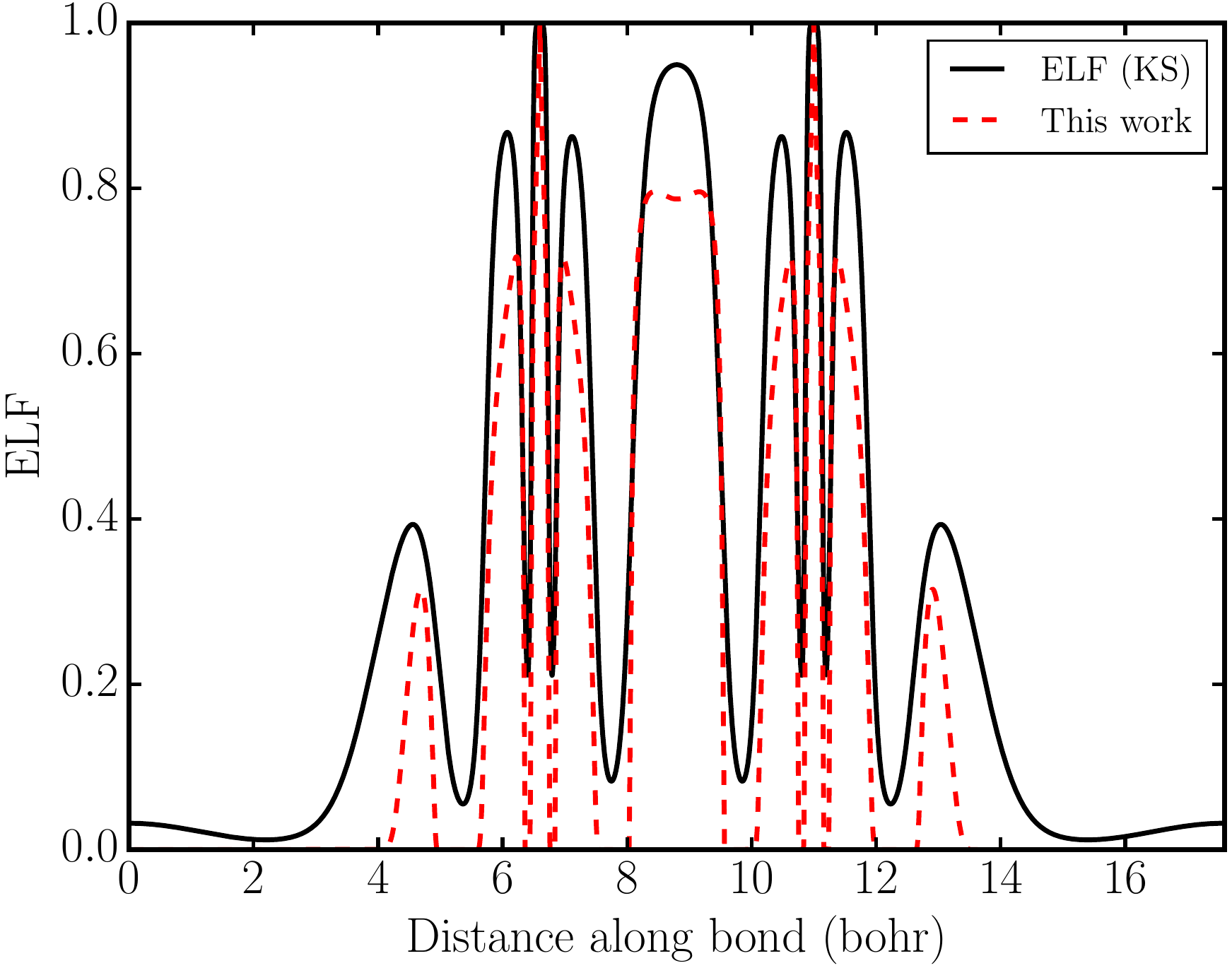}  
   	\caption{(color online) The exact KS DFT ELF (solid, black) and the ELF parameterization in this work (dashed, red) in solid diamond Si along a nearest-neighbor line between the two Si in the unit cell.} 
 	\label{fig:fig_7} 
\end{figure}

\begin{figure}[ht] 
	\centering 
 	\includegraphics[width=0.48\textwidth]{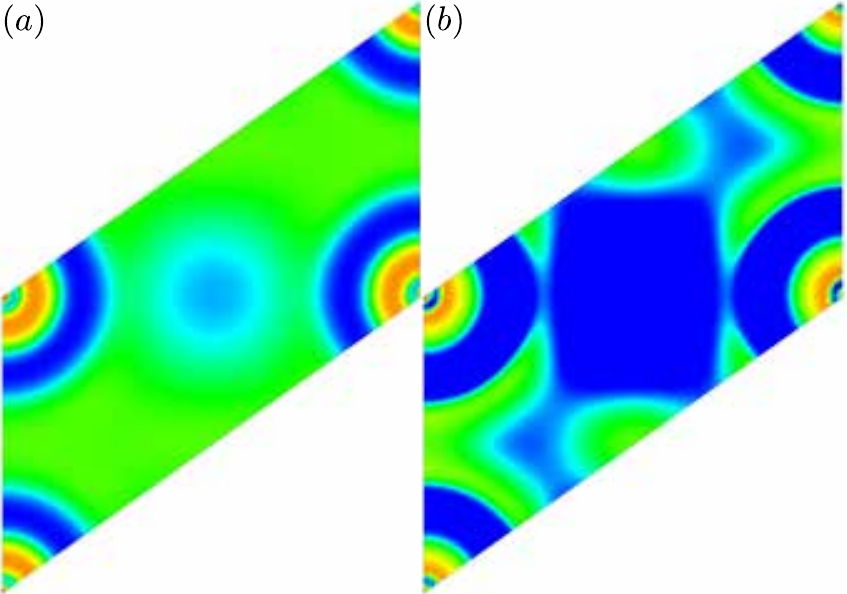} 
  	\caption{Topological surfaces of ELF for bulk Al in the fcc structure. (a) ELF based on self-consistent KS orbitals. (b) ELF based on the parameterisation of this work.} 
 	\label{fig:fig_8} 
\end{figure}

\begin{figure} 
	\centering 
 	\includegraphics[width=0.48\textwidth]{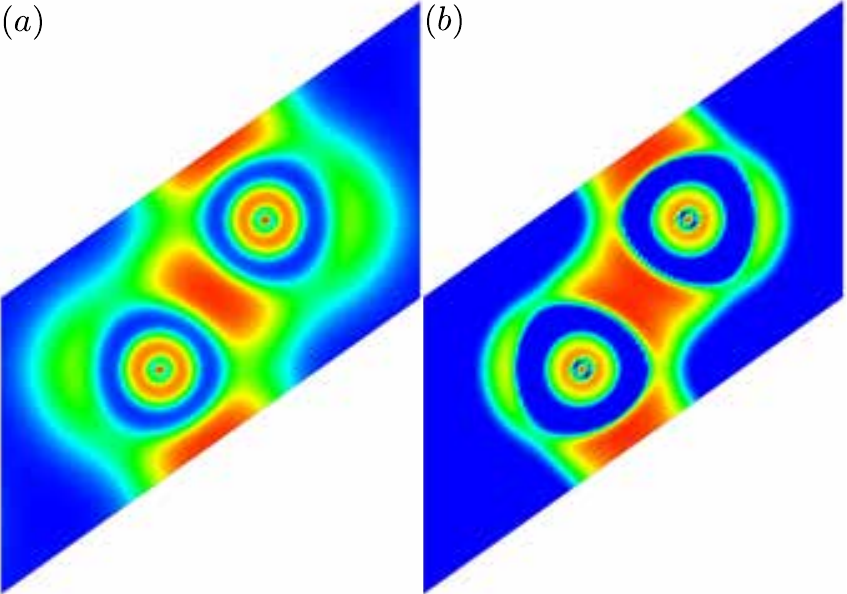} 
  	\caption{Topological surfaces of ELF for bulk Si in the diamond structure. (a) ELF based on self-consistent KS orbitals. (b) ELF based on the parameterisation of this work. } 
 	\label{fig:fig_9} 
\end{figure}

A direct comparison of the parameterization and the real ELF for Si and Al is shown along lines of atomic bonds in Figs.~(\ref{fig:fig_6}) and (\ref{fig:fig_7}), and in 2D planes in Figs.~(\ref{fig:fig_8}) and (\ref{fig:fig_9}). The parameterized ELF visually mimics the qualitative behavior of the real ELF. The two quantities display similar structure in the atomic shell and core regions. The interatomic bonding regions are also visually similar, in particular for the  $sp$-hybridized bonds in Si. However, the parameterization is not a quantitatively accurate approximation of ELF in the studied systems. The height of the maximums are too low in the outer shells. In some regions of moderate to high ELF values, the slope of the parameterization is incorrect. For example, the middle of the inter-bonding region in Si has the parameterization show a marked downward slope. Furthermore, in regions where $s^2-q$ are negative, the parameterization is sharply cutting off to exactly zero. In the 2D plot for Al in Fig.~(\ref{fig:fig_8}), there is a corresponding larger level of visual discrepancy in the ``void'' region away from the atoms. In this region the real ELF is moderately high (ca 0.5), but the parameterization is nearly zero.

The qualitative visual similarity suggest that the parameterization in many regions extracts similar information about the characteristics of electron localization as the real ELF. This is particularly clear when comparing the $sp$-hybridized bonds in Si with the metal bonding for Al. Both in the 1D and 2D figures it is clear that the regions with low or negative $s^2-q$ are inaccurately reproduced. This is not so suprising, since the regions of low electron confinement with negative $s^2-q$ do not map onto any HO model system. 

%

%

\begin{figure} 
	\centering 
 	\includegraphics[width=0.48\textwidth]{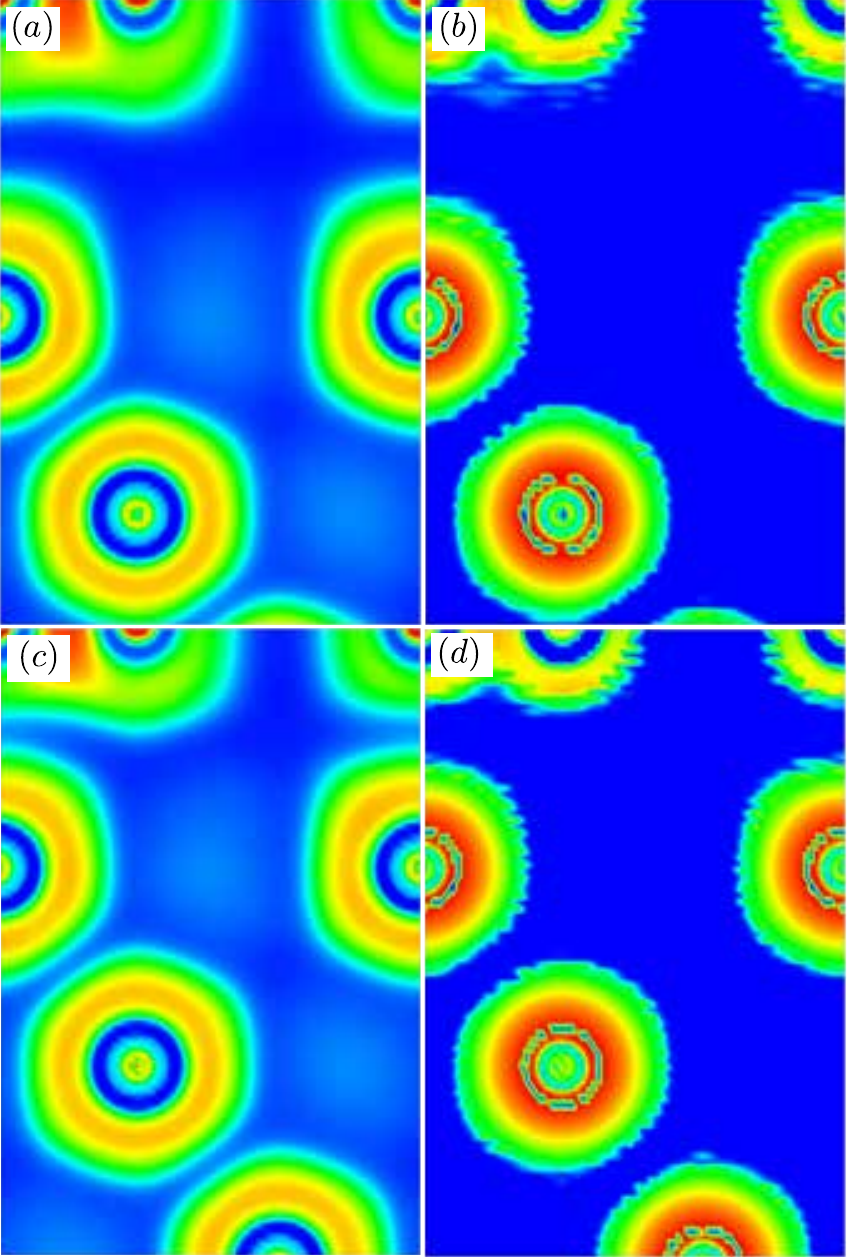} 
  	\caption{Topological surfaces of ELF for graphene on a Ni surface. The topmost atoms are the C atoms in the graphene layer, with the Ni atoms below. The subfigures show (a) ELF based on self-consistent KS orbitals for the case of physisorption. (b) ELF based on the parameterization of this work for the case of physisorption. (c) ELF based on self-consistent KS orbitals for the case of chemisorption. (d) ELF based on the parameterization of this work for the case of chemisorption.} 
 	\label{fig:fig_10} 
\end{figure} 

\subsection{Graphene on Ni}

We now move on to study the real and parameterized ELF in a different context by considering the system graphene on a Ni surface. The use of this system was inspired by the work by Jarvis \emph{et al.} in Ref.~\onlinecite{acs.jpcc.5b08350}, in which ELF was utilized as an indicator of the bonding character as being either physisorbed or chemisorbed. The geometries of the system were chosen to be the same as in Ref.~\onlinecite{PhysRevLett.111.106401}. Moreover, the graphene layer is either physisorbed or chemisorbed depending on its distance to the Ni surface \cite{PhysRevLett.111.106401}. 

The features of the real and parameterized ELF for the two types of bonding physics are shown in the subpanels of Fig.~(\ref{fig:fig_10}). There is overall a visiual simularity for the corrsponding pictures between the parameterized and the real ELF. However, there are also some important differences which are discussed in the following.

For the real ELF, the subpanel Fig.~(\ref{fig:fig_10}a) for the physisorbed case shows the value along the shortest distance between the C and Ni atoms to drop to zero, indicating that there is little to no electron localization here, i.e., no chemical bond. On the other hand, in the chemisorbed bonding situation of Fig.~(\ref{fig:fig_10}c), ELF drops to a low, but non-zero, value. In fact, for this case there is a visual similarity between the Ni-Ni bonds and the Ni-C bonds. Furthermore, the C outer shell takes a different shape between the two bonding situations: in the physisorbed case it appears more elongated towards the Ni surface than in the chemisorbed case.

For the parameterized ELF, the perhaps most striking visual difference to the real ELF is the inner core regions of the Ni atoms. However, the significant discrepancy in absolute value for the maximums of the peaks for the atomic shells was already seen in the 1D plots for Al and Si in Figs.~(\ref{fig:fig_6}) and (\ref{fig:fig_7}). For the 2D surface of the Ni atoms, the parameterized ELF shows a highly osscilating behavior associated with the shells, but the difference in absolute values, along with the larger number of core states in Ni (which are not fully resolved in the 2D plots), combine into a noticeable visual discrepancy to the real ELF.

An arguably more important observation of the difference between the parameterized and real ELF in this system is that the former does not reproduce the distinction between the physisorbed and chemisorbed bonding between the surface and the graphene layer. The parameterized ELF drops off to very nearly zero in both cases. The parameterized ELF also does not show a noticable difference in the shape of the outer shell for C. 

As a general observation, the chemisorbed bonding of graphene on an Ni surface apparently has a relatively weak level of electron localization, since the real ELF has a low value in this region. This places the $s$ and $q$ values in a domain that does not map well, or not at all, onto the HO model system which limits the accuracy of the parameterized ELF. This is the reason the distinction between the chemisorbed and physisorbed case is lost. However, also for the real ELF, the visual distinction is not overly clear compared to, e.g., the covalent $sp$ bonds in Si. The visual distinction between the physisorbed case in Fig.~(\ref{fig:fig_10}) and the $sp$-bond in Fig.~(\ref{fig:fig_9}) is clear in both the parameterized and real ELF. The ability for the parameterized ELF to distinguish these bond types could be demonstrated with greater clarity if we had a system where two different geometries yielded chemisorption with stronger covalent bonds and physisorbtion-type bonding. The authors have so far been unable to find a good such test case. 

\begin{figure}[htb] 
	\centering 
 	\includegraphics[width=\linewidth]{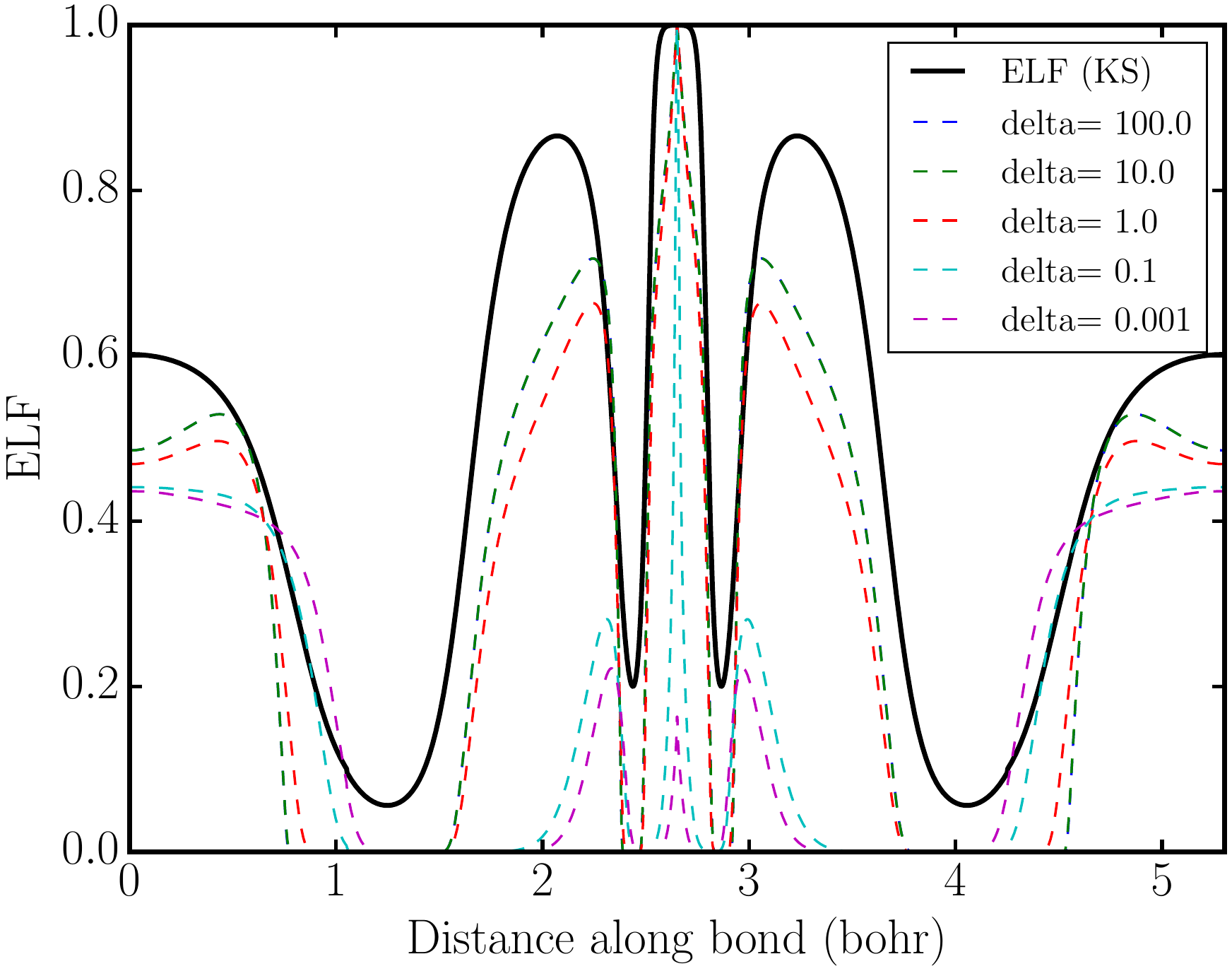} 
   	\caption{(color online) The ELF parameterization in this work for various values of $\delta$ in solid bulk fcc Al along a nearest-neighbohr line between repetitions of the Al atom.}
 	\label{fig:Aldelta} 
\end{figure}

\begin{figure}[htb] 
	\centering 
 	\includegraphics[width=\linewidth]{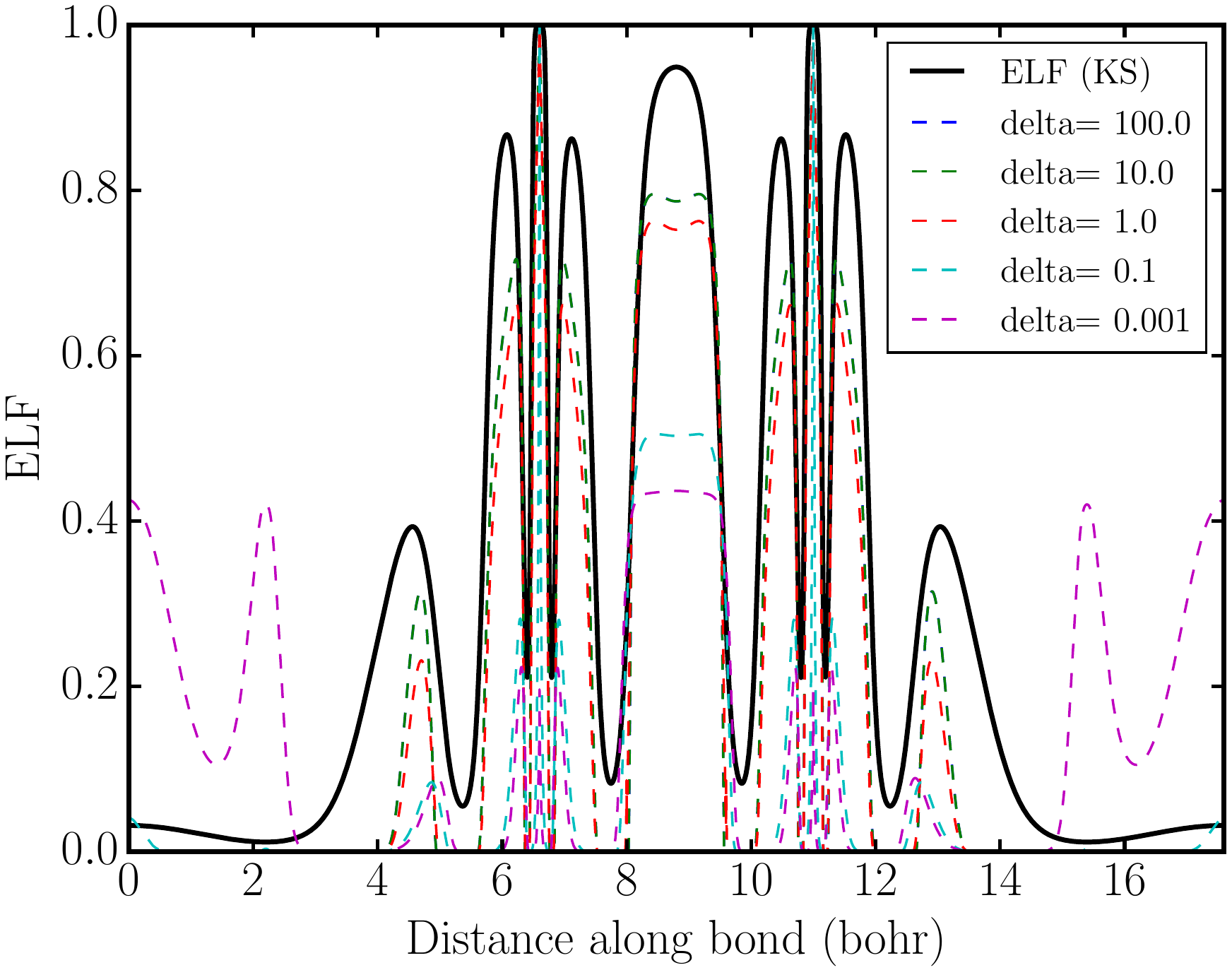} 
   	\caption{(color online) The ELF parameterization in this work for various values of $\delta$ in solid bulk diamond Si along a nearest-neighbohr line between the two Si of the unit cell.}  	
 	\label{fig:Sidelta} 
\end{figure}

Finally, we note that the form of Eq.~(\ref{eq:adj}) may cause one to believe that the sharp cutoff to zero seen in the regions where $s^2-q < 0$ could be reduced by increasing the $\delta$ parameter. However, this is not the case. Figures~(\ref{fig:Aldelta}) and (\ref{fig:Sidelta}) show Eq.~(\ref{eq:finaleq}) for a set of different values of $\delta$, and it is clear that while the size of these regions are altered by $\delta$, the sharp cutoff remains (at least in regions where the similarities with respect to ELF based on KS orbitals are not severely affected by the choice of $\delta$).

%


\section{Summary and Conclusions} 

In this paper we have presented an approximation to the electron localization function given only in terms density-dependent reduced gradients to second order. This allows for a direct way to extract the behavior of localized electronic states such as physical bonds in real systems without the explicit reference to individual orbitals such as those provided by KS theory. Our derivation is based on a parameterized version of the \emph{exact} expression of the non-interacting KS KED in the few-electron limit of a harmonic oscillator model system, from which ELF is then constructed. We then compared the derived model with the exact ELF for simple bulk systems.

Our conclusion is that the derivation gives an expression capable of qualitatively reproducing many of the relevant features of ELF, hence suggesting that it can be used for the visualization of the bonds in solid state and chemistry systems. In our examples, the parameterized ELF clearly distinguishes metallic and dispersive bonds from the stronger covalent $sp$ bonds. Nevertheless, the quantitative agreement is very rough, which may arguably be expected from a parameterization that has no access to an accurate general representation of the KED for any region throughout a general system. Thus, a fundamental limitation of the parameterization is that it can be expected to be accurate only in regions that are in some sense similar to that of the HO model. The HO model with low values of $\alpha$ represents stark electron localization, hence, it stands to reason that regions of high ELF are also the regions where the parameterization is more accurate. For other regions, the expression cuts off to zero rather sharply whereas the real ELF may remain at a low, but non-zero, value. 

Hence, the derivation in this work have produced a quantity that give qualitative pictures of the electron localization that mimics the features of ELF. This is useful in contexts where the electron density is available, but the KS orbitals or the KED are not. Possible applications include analysis of priorly stored data from calculations where these quantities are not included, and the analysis of electronic densities from computational schemes that do not generate KS orbitals---e.g., orbital-free DFT. Visualizations based on the parameterizied ELF have shortcomings in particular for regions where ELF is relatively low, but it appear regions of high and low electron localization are clearly identified. 

Future work may allow extending the parameterization to be more transferable. We note in particular that both the form of the final expression, and our numerical results, suggest that $s^2 -q$ is a relevant indicator of the physics in the regions of a system, and $s^2 -q < 0$ marks regions that are out of reach of the HO model. Furthermore, the quantitative agreement with ELF in regions that are described by the HO model may be improved by taking an additional number of levels into account in the parameterization. Another path forward is to introduce an additional dimension of confinement, which may yield a more transferable expression. Work in these directions is currently underway. 

\begin{acknowledgments} 

R.A.~acknowledges financial support from the Swedish Research Council (VR), Grant No. 2016-04810 and the Swedish e-Science Research center (SeRC). Some of the calculations were performed utilizing the computational resources provided by the Swedish National Infrastructure for Computing (SNIC). A.E.M.~acknowledges partial support from the the Laboratory Directed Research and Development Program at Sandia National Laboratories. Sandia National Laboratories is a multimission laboratory managed and operated by National Technology \& Engineering Solutions of Sandia, LLC, a wholly owned subsidiary of Honeywell International Inc., for the U.S. Department of Energy's National Nuclear Security Administration under contract DE-NA0003525. This work was supported in part by Advanced Simulation and Computing, Physics and Engineering Models, at Los Alamos National Laboratory. Los Alamos National Laboratory, an affirmative action/equal opportunity employer, is managed by Triad National Security, LLC, for the National Nuclear Security Administration of the U.S. Department of Energy under contract 89233218CNA000001. We thank Lars Nordstr\"om, Olga Matthies, and Michael Fechner for useful input on achieving well-converged results when using the electronic structure code Elk. In addition, we acknowledge Andreas Larsson for the suggestion to investigate a system where physisorption and chemisorption could be compared, as well as for useful discussions about the features of ELF in such systems. 

\end{acknowledgments} 

\appendix

\section{Computational Details}\label{sec:comp_details} 

For the least square fitting procedure in order to determine the parameters $\beta$ and $\gamma$ defined by Eq.~(\ref{eq:finaleq}), we use a grid defined by the set of values $\alpha \in (10^{-5},2]$ and $\bar{z} \in [-3,3]$ with a step size of $10^{-3}$ in each direction, after which the numerical optimization is performed by the software \textsc{Mathematica} \cite{mathematica}. 

The Al and Si calculations are done with the all-electron, full-potential linearized-augmented plane wave \cite{lapw} (LAPW) electronic structure code Elk \cite{code_elk} with the local spin-density approximation  \cite{perdew_accurate_1992}. The basis set cutoff (defined as the maximum length of the vectors $\mathbf{G} + \mathbf{k}$) is set to $7.0\ \textrm{bohr}^{-1}$. The $k$-point mesh is $4 \times 4 \times 4$ for Si and $10 \times 10 \times 10$ for Al, with an offset of $(0.5, 0.5, 0.5)$ lattice vectors from the Brillouin zone (BZ) center \cite{chadi_special_1973, monkhorst_special_1976}. A finer radial mesh is used for the muffin-tin functions (the ELK \texttt{lradstp} flag is set to $1$.) The calculations are done without any relativistic correction (the ELK \texttt{solscf} flag is set to $10^7$) to avoid the true KS DFT values in the core regions to be distorted by this correction. 

For the graphene on Ni calculations, the initial structures are relaxed using the Vienna Ab-inito simulation package \cite{PhysRevB.48.13115, PhysRevB.54.11169,PhysRevB.59.1758}, using the local spin-density approximation \cite{perdew_accurate_1992}. The relaxed geometries are imported into Elk to calculate the all-electron densities and KS orbitals. The muffin-tin radii were adjusted to reduce their mismatch since that may lead to numerical issues. Hence, the muffin-tin radius of C from $1.6$ $\textrm{bohr}$ to $1.33$ $\textrm{bohr}$, and for Ni from $2.4$ $\textrm{bohr}$ to $1.9$ $\textrm{bohr}$. The $3s$-state is moved to the interstitial region with a linearization energy of $-3.5$ Hartree. For both species, the matching derivates at the boundary are increased to the third order. Due to the large amounts of vacuum in the system in both cases (the chemisorbed and physisorbed), density sloshing is prevented by adjusting the Broyden mixing parameters \cite{Srivastava_1984}. For the chemisorbed case, the adaptive mixing parameter is set to $0.01$, and for the physisorbed, $0.0004$. The maximum mixing parameter is set to $0.06$ ($0.0008$). The number of empty bands is increased to $8$ and the automatic linearization algorithm is activated. An initial density is found by starting with a rgkmax of $5$, a gmaxvr of $16$, and a lmaxapw of $8$. The parameters are then increased systematically, taking on the the values 8, 11, and 22, and it is observed that the calculated quantities converge. The parameters cannot be increased further, since that results in a divergence of the first and second density derivates in the interstitial region. 


\begin{thebibliography}{44}%
\makeatletter
\providecommand \@ifxundefined [1]{%
 \@ifx{#1\undefined}
}%
\providecommand \@ifnum [1]{%
 \ifnum #1\expandafter \@firstoftwo
 \else \expandafter \@secondoftwo
 \fi
}%
\providecommand \@ifx [1]{%
 \ifx #1\expandafter \@firstoftwo
 \else \expandafter \@secondoftwo
 \fi
}%
\providecommand \natexlab [1]{#1}%
\providecommand \enquote  [1]{``#1''}%
\providecommand \bibnamefont  [1]{#1}%
\providecommand \bibfnamefont [1]{#1}%
\providecommand \citenamefont [1]{#1}%
\providecommand \href@noop [0]{\@secondoftwo}%
\providecommand \href [0]{\begingroup \@sanitize@url \@href}%
\providecommand \@href[1]{\@@startlink{#1}\@@href}%
\providecommand \@@href[1]{\endgroup#1\@@endlink}%
\providecommand \@sanitize@url [0]{\catcode `\\12\catcode `\$12\catcode
  `\&12\catcode `\#12\catcode `\^12\catcode `\_12\catcode `\%12\relax}%
\providecommand \@@startlink[1]{}%
\providecommand \@@endlink[0]{}%
\providecommand \url  [0]{\begingroup\@sanitize@url \@url }%
\providecommand \@url [1]{\endgroup\@href {#1}{\urlprefix }}%
\providecommand \urlprefix  [0]{URL }%
\providecommand \Eprint [0]{\href }%
\providecommand \doibase [0]{http://dx.doi.org/}%
\providecommand \selectlanguage [0]{\@gobble}%
\providecommand \bibinfo  [0]{\@secondoftwo}%
\providecommand \bibfield  [0]{\@secondoftwo}%
\providecommand \translation [1]{[#1]}%
\providecommand \BibitemOpen [0]{}%
\providecommand \bibitemStop [0]{}%
\providecommand \bibitemNoStop [0]{.\EOS\space}%
\providecommand \EOS [0]{\spacefactor3000\relax}%
\providecommand \BibitemShut  [1]{\csname bibitem#1\endcsname}%
\let\auto@bib@innerbib\@empty
\bibitem [{\citenamefont {L{\"o}wdin}(1958)}]{lowdin_2}%
  \BibitemOpen
  \bibfield  {author} {\bibinfo {author} {\bibfnamefont {P.-O.}\ \bibnamefont
  {L{\"o}wdin}},\ }\href {\doibase 10.1002/9780470143483.ch7} {\bibfield
  {journal} {\bibinfo  {journal} {Advances in Chemical Physics}\ }\textbf
  {\bibinfo {volume} {2}},\ \bibinfo {pages} {207} (\bibinfo {year}
  {1958})}\BibitemShut {NoStop}%
\bibitem [{\citenamefont {Dobson}(1991)}]{doi:10.1063/1.460619}%
  \BibitemOpen
  \bibfield  {author} {\bibinfo {author} {\bibfnamefont {J.~F.}\ \bibnamefont
  {Dobson}},\ }\href {\doibase 10.1063/1.460619} {\bibfield  {journal}
  {\bibinfo  {journal} {The Journal of Chemical Physics}\ }\textbf {\bibinfo
  {volume} {94}},\ \bibinfo {pages} {4328} (\bibinfo {year} {1991})},\ \Eprint
  {http://arxiv.org/abs/https://doi.org/10.1063/1.460619}
  {https://doi.org/10.1063/1.460619} \BibitemShut {NoStop}%
\bibitem [{\citenamefont {Pauling}(1948)}]{pauling_bond}%
  \BibitemOpen
  \bibfield  {author} {\bibinfo {author} {\bibfnamefont {L.}~\bibnamefont
  {Pauling}},\ }\href@noop {} {\emph {\bibinfo {title} {The Nature of the
  Chemical Bond}}}\ (\bibinfo  {publisher} {Cornell University Press},\
  \bibinfo {year} {1948})\BibitemShut {NoStop}%
\bibitem [{\citenamefont {Lewis}(1966)}]{lewis}%
  \BibitemOpen
  \bibfield  {author} {\bibinfo {author} {\bibfnamefont {G.~N.}\ \bibnamefont
  {Lewis}},\ }\href@noop {} {\emph {\bibinfo {title} {Valence and Structure of
  Atoms and Molecules}}}\ (\bibinfo  {publisher} {Dover},\ \bibinfo {address}
  {New York},\ \bibinfo {year} {1966})\BibitemShut {NoStop}%
\bibitem [{\citenamefont {Hodgson}\ \emph {et~al.}(2014)\citenamefont
  {Hodgson}, \citenamefont {Ramsden}, \citenamefont {Durrant},\ and\
  \citenamefont {Godby}}]{PhysRevB.90.241107}%
  \BibitemOpen
  \bibfield  {author} {\bibinfo {author} {\bibfnamefont {M.~J.~P.}\
  \bibnamefont {Hodgson}}, \bibinfo {author} {\bibfnamefont {J.~D.}\
  \bibnamefont {Ramsden}}, \bibinfo {author} {\bibfnamefont {T.~R.}\
  \bibnamefont {Durrant}}, \ and\ \bibinfo {author} {\bibfnamefont {R.~W.}\
  \bibnamefont {Godby}},\ }\href {\doibase 10.1103/PhysRevB.90.241107}
  {\bibfield  {journal} {\bibinfo  {journal} {Phys. Rev. B}\ }\textbf {\bibinfo
  {volume} {90}},\ \bibinfo {pages} {241107} (\bibinfo {year}
  {2014})}\BibitemShut {NoStop}%
\bibitem [{\citenamefont {R\"as\"anen}\ \emph {et~al.}(2008)\citenamefont
  {R\"as\"anen}, \citenamefont {Castro},\ and\ \citenamefont
  {Gross}}]{PhysRevB.77.115108}%
  \BibitemOpen
  \bibfield  {author} {\bibinfo {author} {\bibfnamefont {E.}~\bibnamefont
  {R\"as\"anen}}, \bibinfo {author} {\bibfnamefont {A.}~\bibnamefont {Castro}},
  \ and\ \bibinfo {author} {\bibfnamefont {E.~K.~U.}\ \bibnamefont {Gross}},\
  }\href {\doibase 10.1103/PhysRevB.77.115108} {\bibfield  {journal} {\bibinfo
  {journal} {Phys. Rev. B}\ }\textbf {\bibinfo {volume} {77}},\ \bibinfo
  {pages} {115108} (\bibinfo {year} {2008})}\BibitemShut {NoStop}%
\bibitem [{\citenamefont {Burdett}\ and\ \citenamefont
  {McCormick}(1998)}]{doi:10.1021/jp9820774}%
  \BibitemOpen
  \bibfield  {author} {\bibinfo {author} {\bibfnamefont {J.~K.}\ \bibnamefont
  {Burdett}}\ and\ \bibinfo {author} {\bibfnamefont {T.~A.}\ \bibnamefont
  {McCormick}},\ }\href {\doibase 10.1021/jp9820774} {\bibfield  {journal}
  {\bibinfo  {journal} {The Journal of Physical Chemistry A}\ }\textbf
  {\bibinfo {volume} {102}},\ \bibinfo {pages} {6366} (\bibinfo {year}
  {1998})},\ \Eprint {http://arxiv.org/abs/https://doi.org/10.1021/jp9820774}
  {https://doi.org/10.1021/jp9820774} \BibitemShut {NoStop}%
\bibitem [{\citenamefont {Becke}\ and\ \citenamefont
  {Edgecombe}(1990)}]{doi:10.1063/1.458517}%
  \BibitemOpen
  \bibfield  {author} {\bibinfo {author} {\bibfnamefont {A.~D.}\ \bibnamefont
  {Becke}}\ and\ \bibinfo {author} {\bibfnamefont {K.~E.}\ \bibnamefont
  {Edgecombe}},\ }\href {\doibase 10.1063/1.458517} {\bibfield  {journal}
  {\bibinfo  {journal} {The Journal of Chemical Physics}\ }\textbf {\bibinfo
  {volume} {92}},\ \bibinfo {pages} {5397} (\bibinfo {year} {1990})},\ \Eprint
  {http://arxiv.org/abs/https://doi.org/10.1063/1.458517}
  {https://doi.org/10.1063/1.458517} \BibitemShut {NoStop}%
\bibitem [{\citenamefont {Savin}\ \emph {et~al.}(1992)\citenamefont {Savin},
  \citenamefont {Jepsen}, \citenamefont {Flad}, \citenamefont {Andersen},
  \citenamefont {Preuss},\ and\ \citenamefont {von
  Schnering}}]{doi:10.1002/anie.199201871}%
  \BibitemOpen
  \bibfield  {author} {\bibinfo {author} {\bibfnamefont {A.}~\bibnamefont
  {Savin}}, \bibinfo {author} {\bibfnamefont {O.}~\bibnamefont {Jepsen}},
  \bibinfo {author} {\bibfnamefont {J.}~\bibnamefont {Flad}}, \bibinfo {author}
  {\bibfnamefont {O.~K.}\ \bibnamefont {Andersen}}, \bibinfo {author}
  {\bibfnamefont {H.}~\bibnamefont {Preuss}}, \ and\ \bibinfo {author}
  {\bibfnamefont {H.~G.}\ \bibnamefont {von Schnering}},\ }\href {\doibase
  10.1002/anie.199201871} {\bibfield  {journal} {\bibinfo  {journal}
  {Angewandte Chemie International Edition in English}\ }\textbf {\bibinfo
  {volume} {31}},\ \bibinfo {pages} {187} (\bibinfo {year} {1992})}\BibitemShut
  {NoStop}%
\bibitem [{\citenamefont {Silvi}\ and\ \citenamefont
  {Savin}(1994)}]{Silvi:1994aa}%
  \BibitemOpen
  \bibfield  {author} {\bibinfo {author} {\bibfnamefont {B.}~\bibnamefont
  {Silvi}}\ and\ \bibinfo {author} {\bibfnamefont {A.}~\bibnamefont {Savin}},\
  }\href {\doibase 10.1038/371683a0} {\bibfield  {journal} {\bibinfo  {journal}
  {Nature}\ }\textbf {\bibinfo {volume} {371}},\ \bibinfo {pages} {683}
  (\bibinfo {year} {1994})}\BibitemShut {NoStop}%
\bibitem [{\citenamefont {Hohenberg}\ and\ \citenamefont
  {Kohn}(1964)}]{hohenberg_inhomogeneous_1964}%
  \BibitemOpen
  \bibfield  {author} {\bibinfo {author} {\bibfnamefont {P.}~\bibnamefont
  {Hohenberg}}\ and\ \bibinfo {author} {\bibfnamefont {W.}~\bibnamefont
  {Kohn}},\ }\href {\doibase 10.1103/PhysRev.136.B864} {\bibfield  {journal}
  {\bibinfo  {journal} {Physical Review}\ }\textbf {\bibinfo {volume} {136}},\
  \bibinfo {pages} {B864} (\bibinfo {year} {1964})}\BibitemShut {NoStop}%
\bibitem [{\citenamefont {Kohn}\ and\ \citenamefont
  {Sham}(1965)}]{kohn_self-consistent_1965}%
  \BibitemOpen
  \bibfield  {author} {\bibinfo {author} {\bibfnamefont {W.}~\bibnamefont
  {Kohn}}\ and\ \bibinfo {author} {\bibfnamefont {L.~J.}\ \bibnamefont
  {Sham}},\ }\href {\doibase 10.1103/PhysRev.140.A1133} {\bibfield  {journal}
  {\bibinfo  {journal} {Physical Review}\ }\textbf {\bibinfo {volume} {140}},\
  \bibinfo {pages} {A1133} (\bibinfo {year} {1965})}\BibitemShut {NoStop}%
\bibitem [{\citenamefont {Fuster}\ \emph {et~al.}(2000)\citenamefont {Fuster},
  \citenamefont {Sevin},\ and\ \citenamefont {Silvi}}]{doi:10.1021/jp992783k}%
  \BibitemOpen
  \bibfield  {author} {\bibinfo {author} {\bibfnamefont {F.}~\bibnamefont
  {Fuster}}, \bibinfo {author} {\bibfnamefont {A.}~\bibnamefont {Sevin}}, \
  and\ \bibinfo {author} {\bibfnamefont {B.}~\bibnamefont {Silvi}},\ }\href
  {\doibase 10.1021/jp992783k} {\bibfield  {journal} {\bibinfo  {journal} {The
  Journal of Physical Chemistry A}\ }\textbf {\bibinfo {volume} {104}},\
  \bibinfo {pages} {852} (\bibinfo {year} {2000})},\ \Eprint
  {http://arxiv.org/abs/https://doi.org/10.1021/jp992783k}
  {https://doi.org/10.1021/jp992783k} \BibitemShut {NoStop}%
\bibitem [{\citenamefont {Savin}\ \emph {et~al.}(1997)\citenamefont {Savin},
  \citenamefont {Nesper}, \citenamefont {Wengert},\ and\ \citenamefont
  {F{\"a}ssler}}]{doi:10.1002/anie.199718081}%
  \BibitemOpen
  \bibfield  {author} {\bibinfo {author} {\bibfnamefont {A.}~\bibnamefont
  {Savin}}, \bibinfo {author} {\bibfnamefont {R.}~\bibnamefont {Nesper}},
  \bibinfo {author} {\bibfnamefont {S.}~\bibnamefont {Wengert}}, \ and\
  \bibinfo {author} {\bibfnamefont {T.~F.}\ \bibnamefont {F{\"a}ssler}},\
  }\href {\doibase 10.1002/anie.199718081} {\bibfield  {journal} {\bibinfo
  {journal} {Angewandte Chemie International Edition in English}\ }\textbf
  {\bibinfo {volume} {36}},\ \bibinfo {pages} {1808} (\bibinfo {year}
  {1997})}\BibitemShut {NoStop}%
\bibitem [{\citenamefont {Gatti}(2005)}]{gatti_2005}%
  \BibitemOpen
  \bibfield  {author} {\bibinfo {author} {\bibfnamefont {C.}~\bibnamefont
  {Gatti}},\ }\href@noop {} {\bibfield  {journal} {\bibinfo  {journal}
  {Zeitschrift f{\"u}r Kristallographie-Crystalline Materials}\ }\textbf
  {\bibinfo {volume} {220}},\ \bibinfo {pages} {399} (\bibinfo {year}
  {2005})}\BibitemShut {NoStop}%
\bibitem [{\citenamefont {Kohout}(2004)}]{doi:10.1002/qua.10768}%
  \BibitemOpen
  \bibfield  {author} {\bibinfo {author} {\bibfnamefont {M.}~\bibnamefont
  {Kohout}},\ }\href {\doibase 10.1002/qua.10768} {\bibfield  {journal}
  {\bibinfo  {journal} {International Journal of Quantum Chemistry}\ }\textbf
  {\bibinfo {volume} {97}},\ \bibinfo {pages} {651} (\bibinfo {year}
  {2004})}\BibitemShut {NoStop}%
\bibitem [{\citenamefont {Parr}\ and\ \citenamefont
  {Yang}(1989)}]{parr_yang_1989}%
  \BibitemOpen
  \bibfield  {author} {\bibinfo {author} {\bibfnamefont {R.~G.}\ \bibnamefont
  {Parr}}\ and\ \bibinfo {author} {\bibfnamefont {W.}~\bibnamefont {Yang}},\
  }\href@noop {} {\emph {\bibinfo {title} {Density Functional Theory of Atoms
  and Molecules}}}\ (\bibinfo  {publisher} {Oxford University Press},\ \bibinfo
  {year} {1989})\BibitemShut {NoStop}%
\bibitem [{\citenamefont {Kresse}\ and\ \citenamefont
  {Hafner}(1993)}]{PhysRevB.48.13115}%
  \BibitemOpen
  \bibfield  {author} {\bibinfo {author} {\bibfnamefont {G.}~\bibnamefont
  {Kresse}}\ and\ \bibinfo {author} {\bibfnamefont {J.}~\bibnamefont
  {Hafner}},\ }\href {\doibase 10.1103/PhysRevB.48.13115} {\bibfield  {journal}
  {\bibinfo  {journal} {Phys. Rev. B}\ }\textbf {\bibinfo {volume} {48}},\
  \bibinfo {pages} {13115} (\bibinfo {year} {1993})}\BibitemShut {NoStop}%
\bibitem [{\citenamefont {Kresse}\ and\ \citenamefont
  {Furthm\"uller}(1996)}]{PhysRevB.54.11169}%
  \BibitemOpen
  \bibfield  {author} {\bibinfo {author} {\bibfnamefont {G.}~\bibnamefont
  {Kresse}}\ and\ \bibinfo {author} {\bibfnamefont {J.}~\bibnamefont
  {Furthm\"uller}},\ }\href {\doibase 10.1103/PhysRevB.54.11169} {\bibfield
  {journal} {\bibinfo  {journal} {Phys. Rev. B}\ }\textbf {\bibinfo {volume}
  {54}},\ \bibinfo {pages} {11169} (\bibinfo {year} {1996})}\BibitemShut
  {NoStop}%
\bibitem [{\citenamefont {Kresse}\ and\ \citenamefont
  {Joubert}(1999)}]{PhysRevB.59.1758}%
  \BibitemOpen
  \bibfield  {author} {\bibinfo {author} {\bibfnamefont {G.}~\bibnamefont
  {Kresse}}\ and\ \bibinfo {author} {\bibfnamefont {D.}~\bibnamefont
  {Joubert}},\ }\href {\doibase 10.1103/PhysRevB.59.1758} {\bibfield  {journal}
  {\bibinfo  {journal} {Phys. Rev. B}\ }\textbf {\bibinfo {volume} {59}},\
  \bibinfo {pages} {1758} (\bibinfo {year} {1999})}\BibitemShut {NoStop}%
\bibitem [{\citenamefont {Bader}(1994)}]{bader1994atoms}%
  \BibitemOpen
  \bibfield  {author} {\bibinfo {author} {\bibfnamefont {R.}~\bibnamefont
  {Bader}},\ }\href {https://books.google.se/books?id=tyVpQgAACAAJ} {\emph
  {\bibinfo {title} {Atoms in Molecules: A Quantum Theory}}},\ International
  Ser. of Monogr. on Chem\ (\bibinfo  {publisher} {Clarendon Press},\ \bibinfo
  {year} {1994})\BibitemShut {NoStop}%
\bibitem [{\citenamefont {Armiento}\ and\ \citenamefont
  {Mattsson}(2002)}]{armiento_subsystem_2002}%
  \BibitemOpen
  \bibfield  {author} {\bibinfo {author} {\bibfnamefont {R.}~\bibnamefont
  {Armiento}}\ and\ \bibinfo {author} {\bibfnamefont {A.~E.}\ \bibnamefont
  {Mattsson}},\ }\href {\doibase 10.1103/PhysRevB.66.165117} {\bibfield
  {journal} {\bibinfo  {journal} {Physical Review B}\ }\textbf {\bibinfo
  {volume} {66}},\ \bibinfo {pages} {165117} (\bibinfo {year}
  {2002})}\BibitemShut {NoStop}%
\bibitem [{\citenamefont {Armiento}\ and\ \citenamefont
  {Mattsson}(2003)}]{armiento_alternative_2003}%
  \BibitemOpen
  \bibfield  {author} {\bibinfo {author} {\bibfnamefont {R.}~\bibnamefont
  {Armiento}}\ and\ \bibinfo {author} {\bibfnamefont {A.~E.}\ \bibnamefont
  {Mattsson}},\ }\href {\doibase 10.1103/PhysRevB.68.245120} {\bibfield
  {journal} {\bibinfo  {journal} {Physical Review B}\ }\textbf {\bibinfo
  {volume} {68}},\ \bibinfo {pages} {245120} (\bibinfo {year}
  {2003})}\BibitemShut {NoStop}%
\bibitem [{\citenamefont {Armiento}\ and\ \citenamefont
  {Mattsson}(2005)}]{armiento_functional_2005}%
  \BibitemOpen
  \bibfield  {author} {\bibinfo {author} {\bibfnamefont {R.}~\bibnamefont
  {Armiento}}\ and\ \bibinfo {author} {\bibfnamefont {A.~E.}\ \bibnamefont
  {Mattsson}},\ }\href {\doibase 10.1103/PhysRevB.72.085108} {\bibfield
  {journal} {\bibinfo  {journal} {Physical Review B}\ }\textbf {\bibinfo
  {volume} {72}},\ \bibinfo {pages} {085108} (\bibinfo {year}
  {2005})}\BibitemShut {NoStop}%
\bibitem [{\citenamefont {Hao}\ \emph {et~al.}(2010)\citenamefont {Hao},
  \citenamefont {Armiento},\ and\ \citenamefont
  {Mattsson}}]{hao_subsystem_2010}%
  \BibitemOpen
  \bibfield  {author} {\bibinfo {author} {\bibfnamefont {F.}~\bibnamefont
  {Hao}}, \bibinfo {author} {\bibfnamefont {R.}~\bibnamefont {Armiento}}, \
  and\ \bibinfo {author} {\bibfnamefont {A.~E.}\ \bibnamefont {Mattsson}},\
  }\href {\doibase 10.1103/PhysRevB.82.115103} {\bibfield  {journal} {\bibinfo
  {journal} {Physical Review B}\ }\textbf {\bibinfo {volume} {82}},\ \bibinfo
  {pages} {115103} (\bibinfo {year} {2010})}\BibitemShut {NoStop}%
\bibitem [{\citenamefont {Mattsson}\ and\ \citenamefont
  {Armiento}(2010)}]{mattsson_armiento_2010}%
  \BibitemOpen
  \bibfield  {author} {\bibinfo {author} {\bibfnamefont {A.~E.}\ \bibnamefont
  {Mattsson}}\ and\ \bibinfo {author} {\bibfnamefont {R.}~\bibnamefont
  {Armiento}},\ }\href@noop {} {\bibfield  {journal} {\bibinfo  {journal}
  {International Journal of Quantum Chemistry}\ }\textbf {\bibinfo {volume}
  {110}},\ \bibinfo {pages} {2274} (\bibinfo {year} {2010})}\BibitemShut
  {NoStop}%
\bibitem [{\citenamefont {Lindmaa}\ \emph {et~al.}(2014)\citenamefont
  {Lindmaa}, \citenamefont {Mattsson},\ and\ \citenamefont
  {Armiento}}]{PhysRevB.90.075139}%
  \BibitemOpen
  \bibfield  {author} {\bibinfo {author} {\bibfnamefont {A.}~\bibnamefont
  {Lindmaa}}, \bibinfo {author} {\bibfnamefont {A.~E.}\ \bibnamefont
  {Mattsson}}, \ and\ \bibinfo {author} {\bibfnamefont {R.}~\bibnamefont
  {Armiento}},\ }\href {\doibase 10.1103/PhysRevB.90.075139} {\bibfield
  {journal} {\bibinfo  {journal} {Phys. Rev. B}\ }\textbf {\bibinfo {volume}
  {90}},\ \bibinfo {pages} {075139} (\bibinfo {year} {2014})}\BibitemShut
  {NoStop}%
\bibitem [{\citenamefont {Hao}\ \emph {et~al.}(2014)\citenamefont {Hao},
  \citenamefont {Armiento},\ and\ \citenamefont
  {Mattsson}}]{doi:10.1063/1.4871738}%
  \BibitemOpen
  \bibfield  {author} {\bibinfo {author} {\bibfnamefont {F.}~\bibnamefont
  {Hao}}, \bibinfo {author} {\bibfnamefont {R.}~\bibnamefont {Armiento}}, \
  and\ \bibinfo {author} {\bibfnamefont {A.~E.}\ \bibnamefont {Mattsson}},\
  }\href {\doibase 10.1063/1.4871738} {\bibfield  {journal} {\bibinfo
  {journal} {The Journal of Chemical Physics}\ }\textbf {\bibinfo {volume}
  {140}},\ \bibinfo {pages} {18A536} (\bibinfo {year} {2014})},\ \Eprint
  {http://arxiv.org/abs/https://doi.org/10.1063/1.4871738}
  {https://doi.org/10.1063/1.4871738} \BibitemShut {NoStop}%
\bibitem [{\citenamefont {Kohn}\ and\ \citenamefont
  {Mattsson}(1998)}]{kohn_edge_1998}%
  \BibitemOpen
  \bibfield  {author} {\bibinfo {author} {\bibfnamefont {W.}~\bibnamefont
  {Kohn}}\ and\ \bibinfo {author} {\bibfnamefont {A.~E.}\ \bibnamefont
  {Mattsson}},\ }\href {\doibase 10.1103/PhysRevLett.81.3487} {\bibfield
  {journal} {\bibinfo  {journal} {Physical Review Letters}\ }\textbf {\bibinfo
  {volume} {81}},\ \bibinfo {pages} {3487} (\bibinfo {year}
  {1998})}\BibitemShut {NoStop}%
\bibitem [{\citenamefont {Fermi}(1927)}]{fermi_metodo_1927}%
  \BibitemOpen
  \bibfield  {author} {\bibinfo {author} {\bibfnamefont {E.}~\bibnamefont
  {Fermi}},\ }\href@noop {} {\ \textbf {\bibinfo {volume} {6}},\ \bibinfo
  {pages} {602} (\bibinfo {year} {1927})}\BibitemShut {NoStop}%
\bibitem [{\citenamefont {Thomas}(1927)}]{thomas_calculation_1927}%
  \BibitemOpen
  \bibfield  {author} {\bibinfo {author} {\bibfnamefont {L.~H.}\ \bibnamefont
  {Thomas}},\ }\href {\doibase 10.1017/S0305004100011683} {\bibfield  {journal}
  {\bibinfo  {journal} {Mathematical Proceedings of the Cambridge Philosophical
  Society}\ }\textbf {\bibinfo {volume} {23}},\ \bibinfo {pages} {542}
  (\bibinfo {year} {1927})}\BibitemShut {NoStop}%
\bibitem [{\citenamefont {Dirac}(1930)}]{dirac_note_1930}%
  \BibitemOpen
  \bibfield  {author} {\bibinfo {author} {\bibfnamefont {P.~A.~M.}\
  \bibnamefont {Dirac}},\ }\href {\doibase 10.1017/S0305004100016108}
  {\bibfield  {journal} {\bibinfo  {journal} {Mathematical Proceedings of the
  Cambridge Philosophical Society}\ }\textbf {\bibinfo {volume} {26}},\
  \bibinfo {pages} {376} (\bibinfo {year} {1930})}\BibitemShut {NoStop}%
\bibitem [{\citenamefont {Perdew}\ and\ \citenamefont
  {Yue}(1986)}]{perdew_accurate_1986}%
  \BibitemOpen
  \bibfield  {author} {\bibinfo {author} {\bibfnamefont {J.~P.}\ \bibnamefont
  {Perdew}}\ and\ \bibinfo {author} {\bibfnamefont {W.}~\bibnamefont {Yue}},\
  }\href {\doibase 10.1103/PhysRevB.33.8800} {\bibfield  {journal} {\bibinfo
  {journal} {Physical Review B}\ }\textbf {\bibinfo {volume} {33}},\ \bibinfo
  {pages} {8800} (\bibinfo {year} {1986})}\BibitemShut {NoStop}%
\bibitem [{\citenamefont {Lieb}\ and\ \citenamefont
  {Simon}(1973)}]{lieb_thomas-fermi_1973}%
  \BibitemOpen
  \bibfield  {author} {\bibinfo {author} {\bibfnamefont {E.~H.}\ \bibnamefont
  {Lieb}}\ and\ \bibinfo {author} {\bibfnamefont {B.}~\bibnamefont {Simon}},\
  }\href {\doibase 10.1103/PhysRevLett.31.681} {\bibfield  {journal} {\bibinfo
  {journal} {Physical Review Letters}\ }\textbf {\bibinfo {volume} {31}},\
  \bibinfo {pages} {681} (\bibinfo {year} {1973})}\BibitemShut {NoStop}%
\bibitem [{\citenamefont {Abramowitz}\ and\ \citenamefont
  {Stegun}(1965)}]{stegun_1965}%
  \BibitemOpen
  \bibfield  {author} {\bibinfo {author} {\bibfnamefont {M.}~\bibnamefont
  {Abramowitz}}\ and\ \bibinfo {author} {\bibfnamefont {I.~A.}\ \bibnamefont
  {Stegun}},\ }\href@noop {} {\emph {\bibinfo {title} {Handbook of Mathematical
  Functions with Formulas, Graphs, and Mathematical Tables}}}\ (\bibinfo
  {publisher} {Dover Books on Mathematics},\ \bibinfo {year}
  {1965})\BibitemShut {NoStop}%
\bibitem [{cod()}]{code_elk}%
  \BibitemOpen
  \href {http://elk.sourceforge.net/} {\enquote {\bibinfo {title} {The {Elk}
  {FP-LAPW} code},}\ }\BibitemShut {NoStop}%
\bibitem [{\citenamefont {Perdew}\ and\ \citenamefont
  {Wang}(1992)}]{perdew_accurate_1992}%
  \BibitemOpen
  \bibfield  {author} {\bibinfo {author} {\bibfnamefont {J.}~\bibnamefont
  {Perdew}}\ and\ \bibinfo {author} {\bibfnamefont {Y.}~\bibnamefont {Wang}},\
  }\href {\doibase 10.1103/PhysRevB.45.13244} {\bibfield  {journal} {\bibinfo
  {journal} {Physical Review B}\ }\textbf {\bibinfo {volume} {45}},\ \bibinfo
  {pages} {13244} (\bibinfo {year} {1992})}\BibitemShut {NoStop}%
\bibitem [{\citenamefont {Jarvis}\ \emph {et~al.}(2015)\citenamefont {Jarvis},
  \citenamefont {Taylor}, \citenamefont {Baran}, \citenamefont {Thompson},
  \citenamefont {Saywell}, \citenamefont {Mangham}, \citenamefont {Champness},
  \citenamefont {Larsson},\ and\ \citenamefont {Moriarty}}]{acs.jpcc.5b08350}%
  \BibitemOpen
  \bibfield  {author} {\bibinfo {author} {\bibfnamefont {S.~P.}\ \bibnamefont
  {Jarvis}}, \bibinfo {author} {\bibfnamefont {S.}~\bibnamefont {Taylor}},
  \bibinfo {author} {\bibfnamefont {J.~D.}\ \bibnamefont {Baran}}, \bibinfo
  {author} {\bibfnamefont {D.}~\bibnamefont {Thompson}}, \bibinfo {author}
  {\bibfnamefont {A.}~\bibnamefont {Saywell}}, \bibinfo {author} {\bibfnamefont
  {B.}~\bibnamefont {Mangham}}, \bibinfo {author} {\bibfnamefont {N.~R.}\
  \bibnamefont {Champness}}, \bibinfo {author} {\bibfnamefont {J.~A.}\
  \bibnamefont {Larsson}}, \ and\ \bibinfo {author} {\bibfnamefont
  {P.}~\bibnamefont {Moriarty}},\ }\href {\doibase 10.1021/acs.jpcc.5b08350}
  {\bibfield  {journal} {\bibinfo  {journal} {The Journal of Physical Chemistry
  C}\ }\textbf {\bibinfo {volume} {119}},\ \bibinfo {pages} {27982} (\bibinfo
  {year} {2015})},\ \Eprint
  {http://arxiv.org/abs/https://doi.org/10.1021/acs.jpcc.5b08350}
  {https://doi.org/10.1021/acs.jpcc.5b08350} \BibitemShut {NoStop}%
\bibitem [{\citenamefont {Sun}\ \emph {et~al.}(2013)\citenamefont {Sun},
  \citenamefont {Xiao}, \citenamefont {Fang}, \citenamefont {Haunschild},
  \citenamefont {Hao}, \citenamefont {Ruzsinszky}, \citenamefont {Csonka},
  \citenamefont {Scuseria},\ and\ \citenamefont
  {Perdew}}]{PhysRevLett.111.106401}%
  \BibitemOpen
  \bibfield  {author} {\bibinfo {author} {\bibfnamefont {J.}~\bibnamefont
  {Sun}}, \bibinfo {author} {\bibfnamefont {B.}~\bibnamefont {Xiao}}, \bibinfo
  {author} {\bibfnamefont {Y.}~\bibnamefont {Fang}}, \bibinfo {author}
  {\bibfnamefont {R.}~\bibnamefont {Haunschild}}, \bibinfo {author}
  {\bibfnamefont {P.}~\bibnamefont {Hao}}, \bibinfo {author} {\bibfnamefont
  {A.}~\bibnamefont {Ruzsinszky}}, \bibinfo {author} {\bibfnamefont {G.~I.}\
  \bibnamefont {Csonka}}, \bibinfo {author} {\bibfnamefont {G.~E.}\
  \bibnamefont {Scuseria}}, \ and\ \bibinfo {author} {\bibfnamefont {J.~P.}\
  \bibnamefont {Perdew}},\ }\href {\doibase 10.1103/PhysRevLett.111.106401}
  {\bibfield  {journal} {\bibinfo  {journal} {Phys. Rev. Lett.}\ }\textbf
  {\bibinfo {volume} {111}},\ \bibinfo {pages} {106401} (\bibinfo {year}
  {2013})}\BibitemShut {NoStop}%
\bibitem [{\citenamefont {Inc.}(2019)}]{mathematica}%
  \BibitemOpen
  \bibfield  {author} {\bibinfo {author} {\bibfnamefont {Wolfram Research}\ \bibnamefont
  {Inc.}},\ }\href@noop {} {\emph {\bibinfo {title} {Mathematica, {V}ersion
  11.3}}}\ (\bibinfo  {publisher} {Champaign, IL},\ \bibinfo {year}
  {2019})\BibitemShut {NoStop}%
\bibitem [{\citenamefont {Singh}(1994)}]{lapw}%
  \BibitemOpen
  \bibfield  {author} {\bibinfo {author} {\bibfnamefont {D.}~\bibnamefont
  {Singh}},\ }\href@noop {} {\emph {\bibinfo {title} {Planewaves,
  Pseudopotentials and the LAPW Method}}}\ (\bibinfo  {publisher} {Kluwer
  Academic, Boston},\ \bibinfo {year} {1994})\BibitemShut {NoStop}%
\bibitem [{\citenamefont {Chadi}\ and\ \citenamefont
  {Cohen}(1973)}]{chadi_special_1973}%
  \BibitemOpen
  \bibfield  {author} {\bibinfo {author} {\bibfnamefont {D.~J.}\ \bibnamefont
  {Chadi}}\ and\ \bibinfo {author} {\bibfnamefont {M.~L.}\ \bibnamefont
  {Cohen}},\ }\href {\doibase 10.1103/PhysRevB.8.5747} {\bibfield  {journal}
  {\bibinfo  {journal} {Physical Review B}\ }\textbf {\bibinfo {volume} {8}},\
  \bibinfo {pages} {5747} (\bibinfo {year} {1973})}\BibitemShut {NoStop}%
\bibitem [{\citenamefont {Monkhorst}\ and\ \citenamefont
  {Pack}(1976)}]{monkhorst_special_1976}%
  \BibitemOpen
  \bibfield  {author} {\bibinfo {author} {\bibfnamefont {H.~J.}\ \bibnamefont
  {Monkhorst}}\ and\ \bibinfo {author} {\bibfnamefont {J.~D.}\ \bibnamefont
  {Pack}},\ }\href {\doibase 10.1103/PhysRevB.13.5188} {\bibfield  {journal}
  {\bibinfo  {journal} {Physical Review B}\ }\textbf {\bibinfo {volume} {13}},\
  \bibinfo {pages} {5188} (\bibinfo {year} {1976})}\BibitemShut {NoStop}%
\bibitem [{\citenamefont {Srivastava}(1984)}]{Srivastava_1984}%
  \BibitemOpen
  \bibfield  {author} {\bibinfo {author} {\bibfnamefont {G.~P.}\ \bibnamefont
  {Srivastava}},\ }\href {\doibase 10.1088/0305-4470/17/13/525} {\bibfield
  {journal} {\bibinfo  {journal} {Journal of Physics A: Mathematical and
  General}\ }\textbf {\bibinfo {volume} {17}},\ \bibinfo {pages} {2737}
  (\bibinfo {year} {1984})}\BibitemShut {NoStop}%
\end{thebibliography}

%

\end{document}